\documentclass[pre,showpacs,twocolumn,preprintnumbers,amsmath,amssymb,superscriptaddress]{revtex4}

\usepackage{amsmath}
\usepackage{mathtools}
\usepackage{graphicx}
\usepackage{dcolumn}
\usepackage{bm}
\usepackage{longtable}
\usepackage{epstopdf}

\begin{document}
\preprint{APS/123-QED}

\title{The Single Big Jump Principle in Physical Modelling}

\author{}
\affiliation{}

\author{Alessandro Vezzani}
\affiliation{IMEM, CNR Parco Area delle Scienze 37/A
	43124 Parma}
\affiliation{Dipartimento di Matematica, Fisica e Informatica Universit\`a degli Studi di
Parma, viale G.P.Usberti 7/A, 43100 Parma, Italy}
\author{Eli Barkai}
\affiliation{Department of Physics, Institute of Nanotechnology and Advanced Materials, Bar-Ilan University, Ramat-Gan, 52900, Israel}
\author{Raffaella Burioni}
\affiliation{Dipartimento di Matematica, Fisica e Informatica Universit\`a degli Studi di
Parma, viale G.P.Usberti 7/A, 43100 Parma, Italy}
\affiliation{INFN, Gruppo Collegato di
Parma, viale G.P. Usberti 7/A, 43100 Parma, Italy}

\date{\today}

\begin{abstract}
The big jump principle is a well established mathematical result for sums of independent and identically distributed random variables extracted from a fat tailed distribution. It states that the tail of the distribution of the sum is the same as the distribution of the largest summand. In practice, it means that when in a stochastic process the relevant quantity is a sum of variables, the mechanism leading to rare events is peculiar: instead of being caused by a set of many small deviations all in the same direction,  one jump, the biggest of the lot, provides the main contribution to the rare large fluctuation. We reformulate and elevate the big jump principle beyond its current status to allow it to deal with correlations, finite cutoffs, continuous paths, memory and quenched disorder.    Doing so we are able to predict rare events using the extended big jump principle in L\'evy walks, in a model of laser cooling,  in a scattering process on a heterogeneous structure and in a class  of  L\'evy walks with memory. We argue that the generalized big jump principle can serve as an excellent guideline  for reliable estimates of risk and probabilities of rare events in many complex processes featuring heavy tailed distributions, ranging from contamination spreading  to active transport in the cell. 

\end{abstract}

\maketitle

\section{Introduction}

The estimation of the probability of rare events in  Mathematics, Physics,  Economics and Geophysics has been investigated for decades 
 in the context of extreme value statistics \cite{Gumbel,Mezard,Holger,West}  and large deviation theory \cite{Hollander,Touchete,Vulp,Maj}.
Rare events, like the crash of a stock market, or the overflow of a river or an earthquake
are clearly important but difficult to predict.  A starting point for this class of problems, in the physical and mathematical literature,
is the analysis of the far tail of the distribution in a basic stochastic process, useful in many modelling frameworks, 
i.e. the position of a random walker. Central limit theorem arguments can be used to predict the shape of the central part of 
a bunch of walkers, but  they do not describe the far tail of the packet,
which is driven by the statistics of rare fluctuations. For a random walker, rare events and the characterization of the tail of the density are extremely important. 
Imagine we model the spreading of some deadly poison
in a medium with a random walk  process. If an agent in the medium is sensitive to the poison, 
one would like to estimate the far tail
of the density of the poisonous  particles. 

We advance here the principle of the {\it single big jump}, which is used to analyze rare events in (roughly speaking)
fat tailed processes. Very generally, consider a process  consisting of random displacements, and our observable is 
the sum of the displacements, namely the position of a random walker in space. 
The big jump principle deals with a situation where the far tail of the density of particles starting from a
common origin  is the same as the distribution of the largest jump in the process. 
This means that one big jump is dominating 
the statistics of the rare events of the sum. Thus, instead of experiencing a set of many small
 displacements all in the same direction, which would lead to a rare large (and exponentially suppressed)
fluctuation, one jump, the biggest of the lot, provides the mechanism
to rare events. 

The big jump effect is believed to be at work in several domains of science, ranging from economics to geophysics.  It has been rigorously proven for the sum of independent identically distributed (IID)  random variables extracted from fat tailed distributions \cite{Chistyakov,Foss,Zachary} and in the presence of specific correlations \cite{Geluk,Clusel1}.  Its extension  to more physical processes
is still far from being understood.  Interestingly, this extension would allow for better estimates of risks and a better forecast of catastrophic events in many complex processes featuring heavy tailed distributions, from earthquakes to biology. The main purpose of this paper is to show that,  after technical and conceptual modifications, we will be able to use the principle
to describe rare events in widely applicable physical models.

Mathematically, the big jump principle  is formulated for a set of $N$ IID random variables
$\{ x_1 , \dots x_N\}$ with  
a common fat tailed  (more precisely, subexponential) distribution and it reads
\cite{Foss,Embrechts}:
 \begin{equation}
 \mbox{Prob} \left(x_1 + \dots x_N>X\right) = \mbox{Prob}
( \mbox{max} \{ x_1 , \dots x_N \} > X)
\label{eq01}
\end{equation}
{\em when $X$ is large}. 
This means that 
the tail of the distribution of the largest summand is
the same as the  tail of the sum and in this sense 
the sum is dominated by a single macroscopic  jump (\cite{Bouchaud,Greg,Maj1,Kutner}). 
An example in the IID domain are L\'evy flights in dimension one,  
which deal with a sum of displacements
drawn from a power law probability density function $\lambda(x_i) \sim {x_i^{-(1 + \alpha)} }$, and in this case
one finds that Eq. (\ref{eq01}) is also a simple power law
proportional to $N$ and to $X^{-\alpha}$.

The case of IID random variables is clearly over simplified in physical modelling. In the context of spreading phenomenon,  one simple reason to the breakdown of the simplified IID version of the principle of big jump 
is  that diffusive behaviour  in its far tails is cutoff by finite speed of 
propagation. Thus the decay like $X^{-\alpha}$, predicted by the principle in its current form, is unphysical in the situations we are familiar with, that is with a finite observation time.
One of our goals is to formulate the principle  in more general terms and then show how the far tail of the sum behaves beyond the IID case, when the finite
speed of the particles couples non trivially space and time. This we do with the widely applicable L\'evy walk model  \cite{zaburdaev,zumofen}.

Secondly, the most common way to treat stochastic processes is with the use of stochastic differential equations, for example Langevin equations. In this case the trajectories are continuous and in fact the jumps are infinitesimal, hence at first glance it might seem impossible to use the principle of big jump.  
However  using a level crossing technique \cite{Eli3}, we are able to reformulate the big jump principle also for continuous Langevin processes, thus extending its scope dramatically. For that aim we use a model of cold atoms diffusing in optical lattices  \cite{Eli3,Eli4,Eli5}. 

Thirdly, the current status of the mathematical theory deals with the case of a sum of IID random variables, as mentioned. Clearly in many 
physical situations we have  complex spatio-temporal correlations and these again demand 
a rethinking of the formulation of the big jump principle \cite{Lucilla1,Lucilla2}.  The case studied here is the L\'evy Lorentz gas, which deals with the motion of  test particles with fixed speed
colliding with a set of scatterers with very heterogeneous spacing \cite{klafter,levyrand}.  Finally, we  will consider a correlated version of L\'evy walks, going beyond its renewal assumption, still showing that the principle works and extending it  to a wide range of physical processes with memory.  

Our approach is based on the splitting of the rare event in two contributions: the first one comes from the typical length scale of the process and amounts to calculating a jump rate
function; the latter deals with the estimate of an
effective probability of performing a big jump, much
larger than the characteristic length of the process. The
effects of correlations are then included in the sum over
all paths that contribute to the big jump. Moreover, while in simpler models the identification of the biggest jump is somehow obvious, for correlated and continuous processes we encounter new challenges.
 In all these cases, we are able to obtain an explicit form of the tail of the distribution driven by rare events. We uncover rich physics in the rare events, in the sense that  while the typical fluctuations are described by the standard tool-box of central limit theorems, the rare events reveal the details of the underlying processes, like the non-analytical behaviours found in the L\'evy Lorentz gas (see details below).
 
Both large deviation theory and the big jump principle  investigate statistics of rare events, beyond the traditional central limit theorems. 
However, here end the similarities, as large deviation theory deals with 
systems with exponentially small fluctuations, i.e.
$P(x) \sim \exp( - N I(x))$, where $N$ 
can be  the number of steps in a simple random walk (the well established
 theory is of course much more general). The main focus there
is therefore on the calculation of the rate function $I(x)$. However,
as discussed by Touchette \cite{Touchete}, when a fat tailed is present, such as for example in the two sided Pareto distribution of the summands $x_i$, 
the rate function is equal to zero, so alternative approaches are needed. 
 Further, our work sheds new light on 
the so called infinite covariant densities and strong anomalous diffusion \cite{Eli1}, as we explain further down.

The paper is organised  as follows.
We start with further discussion of the IID case, and then consider the widely applicable L\'evy walk model.
In Sec. III we tackle the problem of big jump definition for continuous trajectories in a Langevin equation modelling cold atom motion, and in Sec. IV we consider the case of quenched disorder in the L\'evy-Lorentz gas, with an extension to a correlated random walk in Sec. V.
Finally in Sec. VI and VII we present a discussion, a list of open problems and our perspectives.

\section{IID Random Variables, L\'evy walks and the single big jump}
\label{walk}

\subsection{IID Random Variables}
Let us first recall the case of IID random variables, which can be considered the well established starting point for our method.
Consider the sum $R=\sum_{i=1} ^N L_i$  of $N$ IID random variables all drawn from a common long tailed Probability Density Function (PDF) e.g:
\begin{equation}
\lambda(L) = {\alpha (L_0)^\alpha  \over L^{1 + \alpha} } 
\label{lambdaL}
\end{equation}
with $0<L_0<L<\infty$. So in our example, we choose a simple power law for the jump size distribution. Notice that, for IID random variables, the big jump principle holds for the wider class of subexponential functions \cite{Foss,Embrechts}, that includes for example the Weibull distribution of the form $\lambda(L)\sim L^{\alpha-1} \exp(-a |L|^\alpha )$ with $0<\alpha<1$. In this  case, all the moments of the distribution are finite.

According to the  {\it single big jump} principle in Eq. (\ref{eq01}), the sum $R$ can be estimated, for large $R$, by the largest value of the summands, i.e.: $\mbox{Prob} (R > \tilde{R} ) \simeq \mbox{Prob} ( \mbox{Max} \{L_i\}  > \tilde{R} )$ \cite{Foss}.  This can be calculated as follows: 
\begin{eqnarray}
\mbox{Prob} (\mbox{Max}\{L_i\}  > \tilde{R} ) & = & 1- \prod_{i=1}^N \left(1-\int_{\tilde{R}}^\infty \lambda(L_i)dL_i\right) \nonumber \\
&=& 1- \left(1- \int_{\tilde{R}}^\infty \lambda(L)dL \right)^N \nonumber \\
&\simeq& N \int_{\tilde{R}}^\infty \lambda(L)dL
\label{eq03b}
\end{eqnarray}
%
%
%
Then the PDF of $R$ is given by the derivative of Eq. (\ref{eq03b}):
\begin{equation}
\mbox{PDF} (R) \sim N \alpha (L_0)^{\alpha}  R^{- 1 - \alpha}. 
\label{eq04}
\end{equation}
This well known result holds for all $\alpha>0$, including the cases $\alpha>2$, where the sum is attracted to the Gaussian central limit theorem. 
The reason for this is that  Eq. (\ref{eq01}) holds for rare events, namely for large $X$, where the central limit theorem does not hold. 
Notice also that technically the big jump principle is related to the field of extreme value statistics, which deals with the question of  the distribution
of the largest random variable drawn from a set on $N$ random variables \cite{Dean,Fyodorov,Ziff,Godreche}. In extreme value theories,  $N$ is usually taken to be large,
which is not a demand for the principle. In particular, focusing on IID random variables described by Eq. (\ref{lambdaL}),  the maximum value
follows a Frech\'et distribution \cite{Gumbel,Mezard,Holger} when $N \to \infty$, and for large $R$ this decays precisely as a fat tailed power law, as indicated in Eq. (\ref{eq04}).
Other types of relations between sums of random variables, not necessarily of power law type and possibly
correlated summands are treated in \cite{Clusel1,Clusel2}.  

\begin{figure}
	\includegraphics[width=0.45\textwidth]{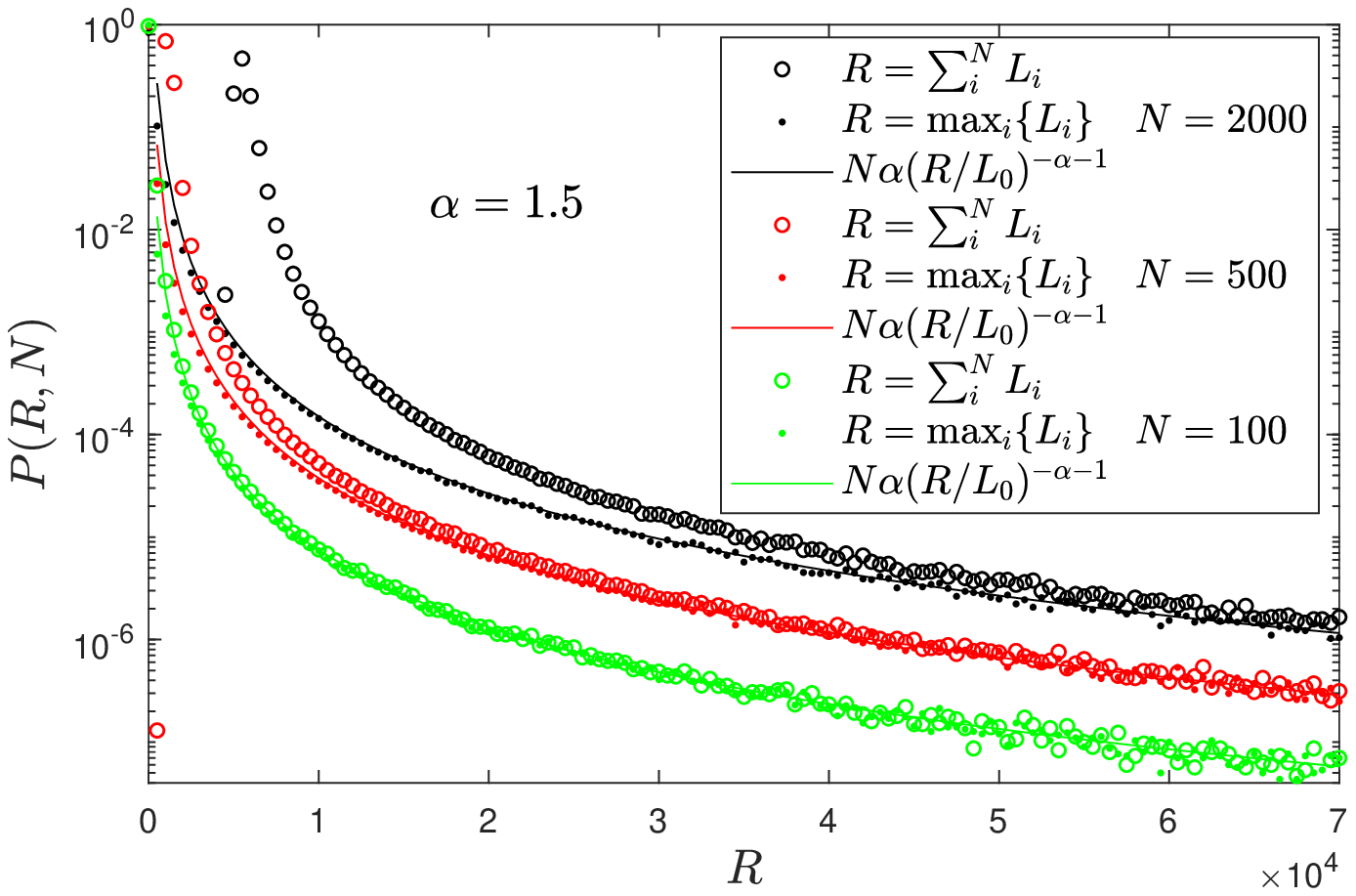}
	\includegraphics[width=0.45\textwidth]{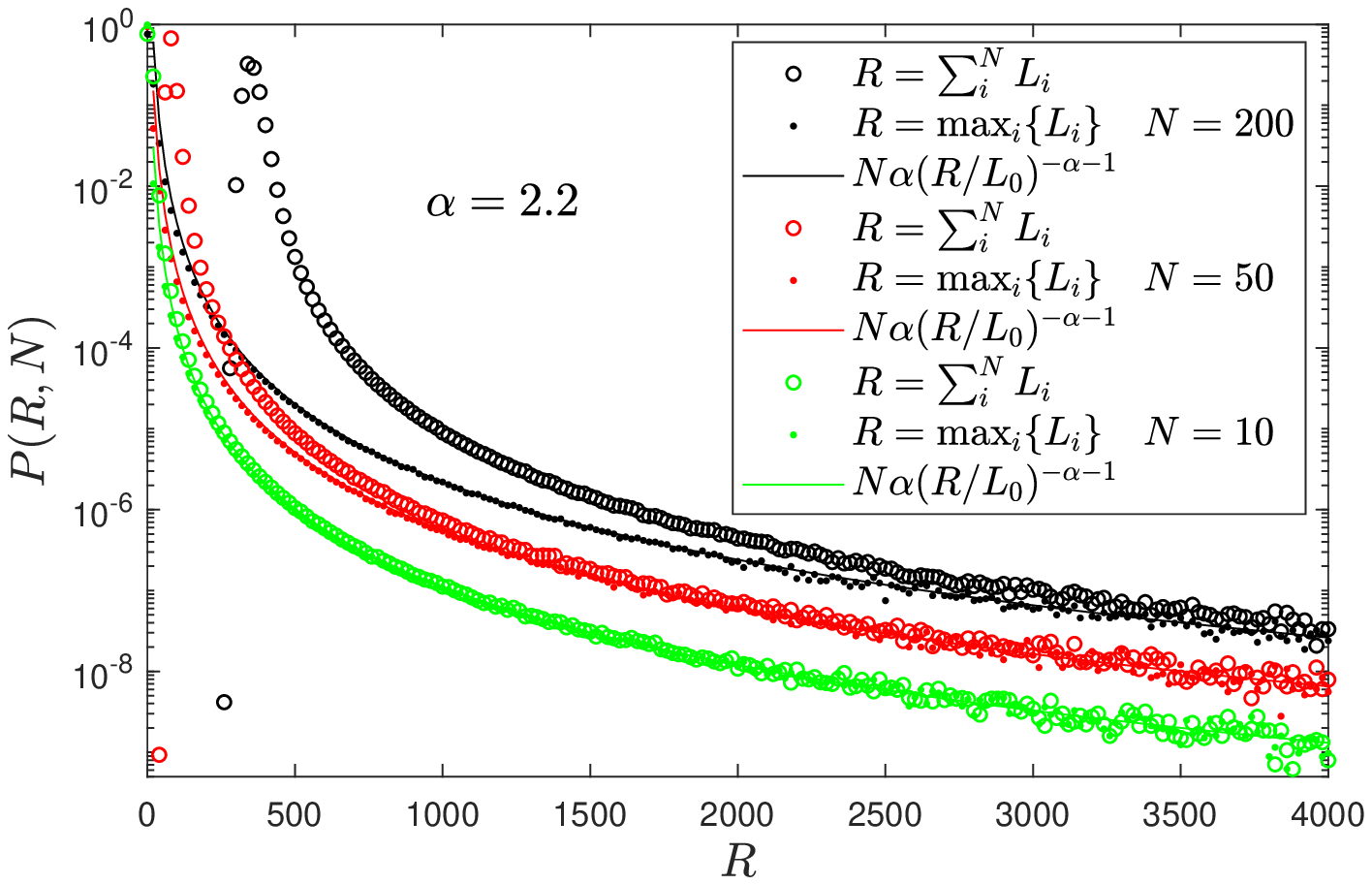}
	\caption{(Color on-line) Numerical verification of the big jump principle: the sum $N$ of IID random variables extracted from (\ref{lambdaL}) is compared to their maximum and to the analytic estimate 
		$\alpha N(R/L_0)^{-\alpha-1}$, here $L_0=1$ . }.
	\label{IIDfig}
\end{figure}

In Fig. \ref{IIDfig} we compare the sum $R$ of $N$ IID random variables extracted from the distribution (\ref{lambdaL}) with their maximum and with the asymptotic estimate in Eq. (\ref{eq04}). The plot shows the efficiency of the single big jump principle: in particular, even at  large $N$, we get a reliable estimate of the asymptotic tail also for values of $R$ which are close to the value that corresponds to the peak of the distribution.

In random walk theory, the sum $R= \sum_{i=1} ^N  L_i$
represents the displacement of the particle starting on the origin, after $N$
steps, and for simplicity we have considered positive random variables, $L_i>0$.
The results are however easily extended to any distribution with power law decay at large $L_i$. 
The case where the PDF of the step $L_i$ is symmetric ($\lambda(L_i) = \lambda(-L_i)$) and $0<\alpha<2$ in the physical literature is called the symmetric L\'evy flight in dimension
one. 
One can argue, at least in the context of random walk theory, 
that the L\'evy flight is marginally physical, as
the mean square displacement of the particle is infinity, $\langle R^2\rangle=\infty$,  
for any $N$.  The unphysical
element of the model is that a long jump takes the same amount of time
as a small jump.  
As mentioned in the introduction, a more physical but still simple model
is the L\'evy walk. Here a velocity is introduced into the model,
so that in a finite time the walker cannot reach arbitrary large distances
and hence the mean square displacement is always finite \cite{zumofen,zaburdaev}. 
The L\'evy walk model has found many applications \cite{zaburdaev}. For example, in the spreading of heat and energy in many body one dimensional  systems, under certain conditions  
the spreading is described by L\'evy laws, which are cut off by sound modes, so that  the speed of sound is a natural cutoff in these systems. 
The far tail of the distribution of the L\'evy walk 
was previously investigated, using the moment generating function approach \cite{Eli1,Eli2,Eli3}.
As we now show, unlike the IID case,  the principle of big jump still holds but it is not completely trivial. 
We will discuss a heuristic derivation of the principle, that will be useful in the following,  and then describe the jump rate method.

\subsection{L\'evy walks}

Let us now consider a one-dimensional L\'evy walk where the length of  the jumps $L_i>0$  is again extracted from the distribution $\lambda(L)$ in Eq. \ref{lambdaL} but in each jump the distance $L_i$ is covered  with probability $1/2$ at velocity $v_i=v$ and with probability $1/2$ at velocity $v_i=-v$ ($v>0$). An event corresponding to the extraction of a new jump 
can be considered as a collision.  The step lengths
and the velocities are mutually independent random variables  and the process is renewed after each jump.
Each step is covered in the finite time $\tau_i=L_i/v$, and one can equivalently define the model by extracting the duration of each step from the distribution $\psi(\tau)=v\lambda(\tau v)$. 
At time $T=0$, the walker begins its motion extracting the first jump, then we observe the system 
at the measurement time $T$. 
The number of collisions $N$ at time $T$ is now given by the condition $\sum_{i=1}^N  \tau_i\leq T < \sum_{i=1} ^{N+1} \tau_i$, so here $N$ is random, unlike the previously studied case of L\'evy flights. The time
$\tau_B= L_B/v = T - \sum_{i=1} ^N \tau_i$ is called
the backward recurrence time. 
The relevant quantity now is the PDF $P(R,T)$ i.e. the probability for the walker to be at time $T$ at distance $R$ from the starting point: $R=|\sum_{i=1} ^N \tau_i v_i + \tau_B v_{i+1}|=
|\sum_{i=1} ^N L_i v_i/v + L_B v_{i+1}/v|$. The process is symmetric with respect to the origin  and therefore the distance $R$ fully describes the density of particles in space. Due to the finite velocity, the walker in a time $T$ cannot reach distances larger than $vT$. Hence we expect that $P(R,T)=0$ for $R>vT$ so that the moments $\langle R^q(T)\rangle$ are clearly finite for any value of $q$.

The L\'evy-Gauss Central Limit Theorem can be used to show that the PDF $P(R,T)$ displays the following scaling form:
$P(R,T)=\ell(T)^{-1}f(R/\ell(T))$ where the scaling length behaves as $\ell(T)\sim T^{1/2} $
for $\alpha>2$, $\ell(T)\sim T^{1/\alpha} $ for $1<\alpha<2$ and as $\ell(T)\sim T$
for $\alpha<1$ \cite{zumofen,zaburdaev}. 
These dynamical  phases are called normal for $\alpha>2$ (since the scaling function $f(.)$ is Gaussian), 
superdiffusive for $1<\alpha<2$ (since the mean square displacement grows faster than normal)
and ballistic for $\alpha<1$. 
The form of the scaling function $f(.)$, the moments of the process and its extensions, for example to higher dimensions, were obtained in previous works \cite{zumofen,zaburdaev,zaburdaev2}. 

The big jump principle does not deal with the scaling of the PDF $P(R,T)$ on the typical length scale $\ell(T)$. The focus here
is on rare events, when $R$ is large and of the order of $vT$. The big jump principle then suggests that,  asymptotically when $\tilde{R}$ is large:
\begin{equation}
\mbox{Prob} \left(R > \tilde{R}\right) = \mbox{Prob}\left( \bar{R}_M> \tilde{R}\right)
\end{equation}
where $\bar{R}_M=\max\{L_1,... L_B\}$. 

Following the derivation for IID presented in the previous Section, the PDF $P(R,T)$ for large $R$ can be estimated as follows.
During a big jump, which is of order of $vT$,
the trajectory does not renew itself in a time interval
$R/v < T$.  In the total remaining time $T - R/v$ the walker is generating attempts
(renewals) to make the big jump. For $\alpha>1$ the average time 
between collision events $\langle \tau \rangle = \int_0 ^\infty d \tau \psi(\tau) {\rm}  \tau$ is finite, and this is the case investigated here. 
The renewal rate is $\langle \tau\rangle^{-1}$ and so the
typical number of renewal is:
\begin{equation}
N_{{\rm eff}} \sim  (T - R/v)/\langle \tau \rangle, .  
\label{neff}
\end{equation}
which provides a nice estimate for large times: $T\gg \langle \tau \rangle$.
We can argue that for $L_0 << \tilde{R}< v T$, $\mbox{Prob} (R> \tilde{R})=
\mbox{Prob} (\bar{R}_M > \tilde{R})$ 
is given by the number of renewals times the probability for a jump to bring 
the particle a distance larger than $\tilde{R}$:
\begin{eqnarray}
\mbox{Prob} (R > \tilde{R} ) &\simeq& N_{{\rm eff}} \int_{\tilde{R}} ^{\infty} \lambda(L ) {\rm d} L \nonumber\\
& = &
{ T - \tilde{R} / v \over \langle \tau \rangle} 
\int_{\tilde{R} } ^\infty \lambda(L) {\rm d} L << 1, 
\label{Plarge}
\end{eqnarray}
while for $\tilde{R} > vT$, $\mbox{Prob} ( R > \tilde{R})= 0 $.
Now we obtain the PDF by taking the derivative. For large $R$ we get
$P(R,T)=0$ if $R> v T$ and 
\begin{eqnarray}
P(R,T) &\simeq& {1 \over v \langle \tau \rangle} \int_{R} ^\infty \lambda(L) {\rm d} L + { T  - R/v \over \langle \tau \rangle} \lambda(R) \nonumber\\
&\simeq& B_0 (R, T) + B_1 (R, T) 
\label{eq05}
\end{eqnarray}
if $R < v T$, with 
\begin{equation}
B_0 (R , T) = {1 \over v \langle \tau \rangle} (L_0)^{\alpha} R^{-\alpha} ,
\end{equation}
and
\begin{equation}
B_1 (R,T) = { T - R/v\over \langle \tau \rangle} \alpha (L_0)^\alpha R^{ - 1 - \alpha} ,
\end{equation}
which is the exact result for the tail of the PDF of the L\'evy walk, obtained by a different method \cite{Eli1,Eli2}.
Compared with the IID case, all we did was to replace $N$ with $N_{{\rm eff}}$,
still this heuristic argument
provides  the known result. This is an indication that
the big jump principle is a useful simple tool to obtain
asymptotic results at ease, and this will be now extended 
to  interpret the physical meaning of the two terms $B_0$ and $B_1$.
This form of the PDF holds for
the scaling region $R\sim vT$, namely for rare events, 
while for $R \sim \ell(T) $ the distribution is described by the L\'evy-Gauss Central Limit Theorem.
So  for $\alpha>1$
the principle of big jump gives the far tail of the distribution and hence is complementary to the Central Limit Theorem. 

Notice that $B_0 (R, T) + B_1 (R, T) \equiv T^{-\alpha} I_\alpha(R/vT)$ in Eq. (\ref{eq05}), with:
\begin{equation}
I_{\alpha} (y) = C
\left[ \alpha y^{-(1+\alpha)} - (\alpha-1) y^{-\alpha}\right],
\label{eq11}
\end{equation}
is not normalized, as its integration diverges due to the pole in $y \to 0$ in Eq. (\ref {eq11}), which stems from $B_1(R,T)$.
This is hardly surprising, since
as we have just mentioned this equation works only for $R \sim vT$ and the divergence stems from the $R \to 0$ limit.
The non-normalized expression $T^{-\alpha} I_\alpha(R/vT)$ is called an {\em infinite density}, since while being non-normalized (hence the term infinite), it can be used to compute exactly the moments $\langle R^q\rangle$ with $q>\alpha$. 
The idea is that moments which are integrable with respect to this non-normalizable function can be computed
as if it was a standard density. In other words, the function
$R^q I_\alpha(R/vT)/T^\alpha$ is integrable, as $R^q$ cures 
the divergence of the density 
on $R \to 0$ when $q> \alpha$. Infinite densities play an important role also 
in ergodic theory. We remark that in $I_\alpha(R/vT)$ there is a scaling 
length that grows linearly with time. This means that, when the single big jump
dominates the dynamics, the typical space-time relation of the single step is ballistic (for any $\alpha$), and this ballistic scale controls the behavior 
of the far tail of the density.

\begin{figure}
	\includegraphics[width=0.45\textwidth]{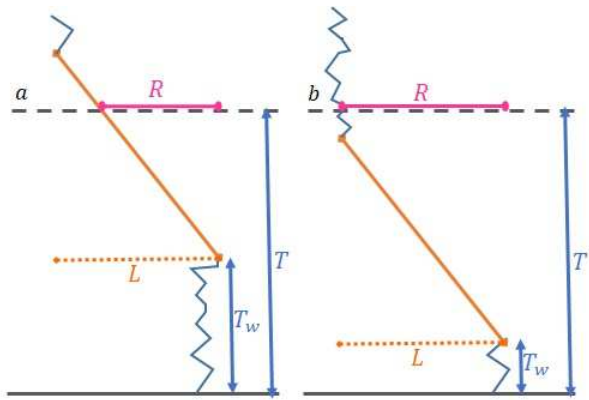}
	\caption{
		The big jump starts at time $T_w$ and this jump can either lead you to the
		time horizon of the L\'evy walk (left panel) or it may start and end before the completion
		of the process at time $T$ (right panel). These processes are used in the text to explain
		the meaning of the two terms contributing to the far tail distribution of the L\'evy walker. 
		To investigate rare fluctuations of the total displacement, we need to consider only the big jump, while in the time interval $(0,T_w)$ we generate attempts to make the big jump with a rate specified in the text.}
	\label{LWfig}
\end{figure}

\subsection{L\'evy walks: the jump rate and the Big Jump}

In L\'evy walks, the mean waiting time between renewals is finite for $\alpha>1$. We will  therefore use now an alternative approach to derive the tail of the PDF $P(R,T)$,
based on the {\em rate of attempts} of making a big jump. 
This will provide a physical interpretation of the terms $B_0 (R,T)$ and $B_1 (R,T)$ and it will be useful when we will apply the big jump principle in more complex processes further on. 

Let us consider $N_{T}$ the average number of jumps up to time $T$, we define the jump rate
$r_{{\rm eff}} (T_w) = dN_T / dT$. For the sake of the analysis performed in the next sections,  $r_{{\rm eff}}$ can depend on time, while here  $r_{{\rm eff}} (T_w) = \langle \tau \rangle ^{-1}$, since the mean duration of a step is  constant and finite for $\alpha>1$. We also define $p_{{\rm tot}}(L,T_w)dL dT_w$, i.e. the probability that the walker in the time interval $[T_w,T_w+dT_w]$ performs a jump of length between $L$ and $L+dL$. Since the jump length $L$ is uncorrelated from the jumping time, it follows that $p_{{\rm tot}}(L,T_w)dL dT_w=  r_{{\rm eff}}(T_w)\cdot \lambda(L) dL dT_w$. 

Now, at large $R$ the big jump principle states that the PDF $P(R,T)$ is determined by the largest jump occurring up to time $T$.  Let us analyze heuristically the motion, following Fig. 2: at a time $T_w\in [0, T]$ the walker takes its longest step $L\sim vT$. The propagation of the walker up to  $T_w$ is of order of $\ell(T_w)\ll L$ and it can be neglected. After this big jump, the motion of the walker can again be neglected since it covers a distance of order $\ell(T-R/v-T_w)\ll L$.  Summing up, in the big jump picture:
the walker remains at the starting point up to time $T_w$ , then it performs a jump of length $L\sim vT$, after that it remains in $L$. We now have to sum $p_{\rm tot}(L,T_w)dL dT_w$ over all the paths ($L$ and $T_w$) that take the walker at a distance $R$ at time $T$ and, as shown in Fig. 2, two different kind of processes are possible. 

In the first path in Fig. 2, $L>v(T-T_w)$, the walker is still moving in the big jump at $T$ and $R=v(T-T_w)$, i.e. $L>R$, $T_w=T-R/v$ and $dT_w=dR/v$. Clearly all the jumps of length $L>R$ contribute to the process ending in $R$, so the probability density $B_0(R,T)$ of this process is:
\begin{eqnarray}
B_0(R,T) dR & = & r_{{\rm eff}}(T-R/v) dR/v \int_{R}^{\infty}\lambda(L) dL \nonumber \\
& = &
\frac{dR}{ v\langle \tau \rangle}  \left(\frac{L_0}{R}\right)^\alpha,
\label{g0w}
\end{eqnarray}
since here $r_{{\rm eff}}(T-R/v) = r_{{\rm eff}}(T_w) = {\langle \tau \rangle^{-1} }$. If $L<v(T-T_w)$, (second path in Fig. 2) the walker ends its motion in $L$ so that $R=L$ and $dR=dL$.  
This process is possible for all the $T_w$ such that $T_w<T-R/v$.
The probability of reaching $R$ is then obtained integrating over the different $T_w$ arriving at the same position: 
\begin{eqnarray}
B_1(R,T) dR &=& \frac{\alpha L_0^\alpha dR}{R^{1+\alpha}} \int_{0}^{T-R/v} r_{{\rm eff}}(T_w) dT_w \nonumber\\
& = & \frac{\alpha L_0^\alpha (T-R/v)}{\langle \tau \rangle R^{1+\alpha}}dR.
\label{g1w}
\end{eqnarray}
Summing Eqs. (\ref{g0w}) and (\ref{g1w}) we get Eq. (\ref{eq05}). This explains that two different processes give rise to the terms $B_0(R,T)$ and $B_1(R,T)$.

The results obtained here can be easily generalized to different definitions
of L\'evy walks. For example in dimensions larger than one, or in the case 
of random velocity  (e.g. Gaussian velocity distribution).
In the wait first model \cite{zaburdaev,waitF,waitF2},
where the particle is localized in space, and then makes 
an instantaneous  displacement, big jumps can be generated only by 
the second process since the particle is at rest at the moment of observation $T$,
and the tail of $P(R,T)$ is given by $B_1$ only. Another possible 
extension where the big jump principle can be applied is the model where the motion of the particle is not ballistic 
\cite{Dentz} . An important example of accelerated motion will be discussed in detail in
Section \ref{atom}.

\begin{figure}
	\includegraphics[width=0.45\textwidth]{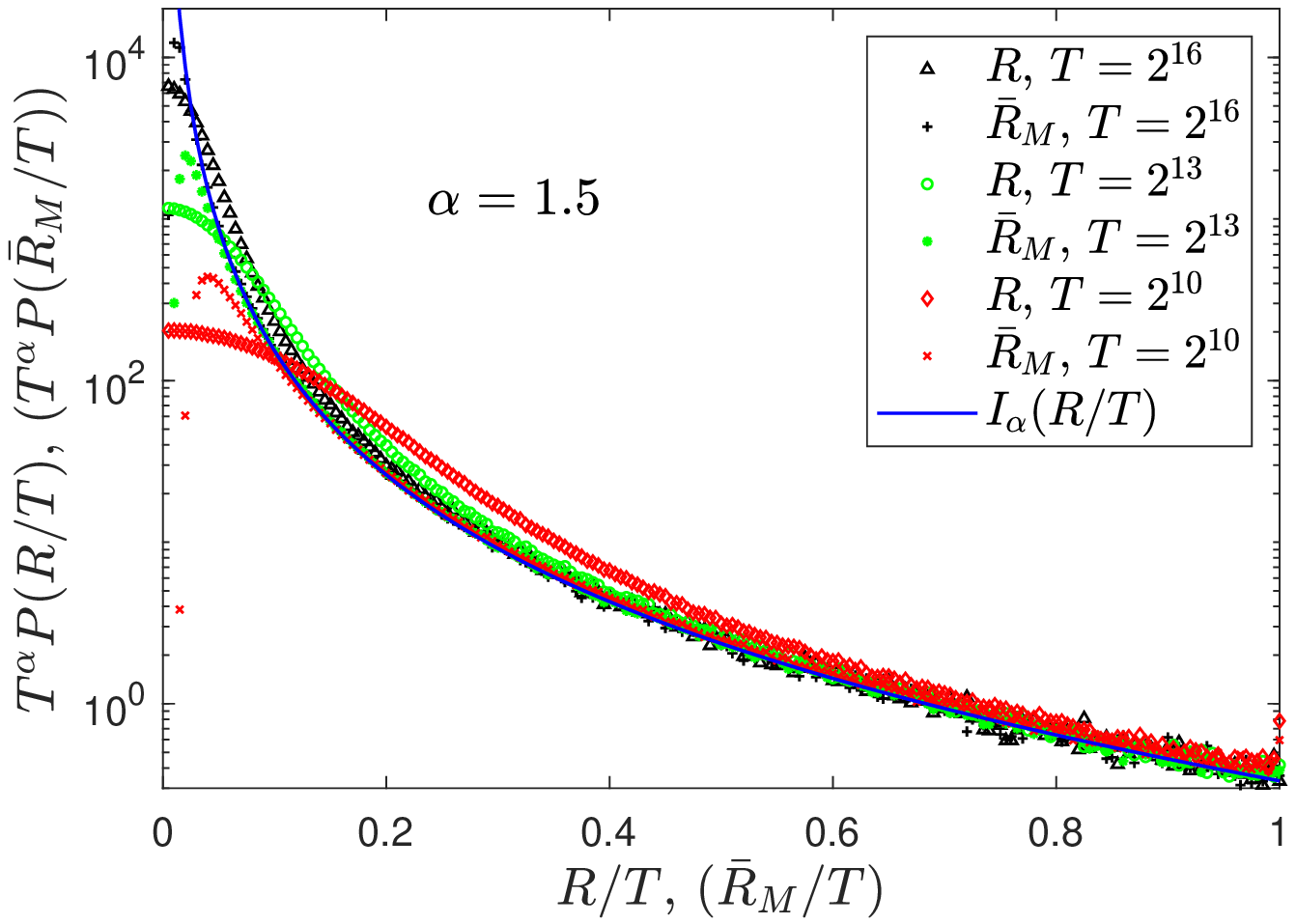}
	\includegraphics[width=0.45\textwidth]{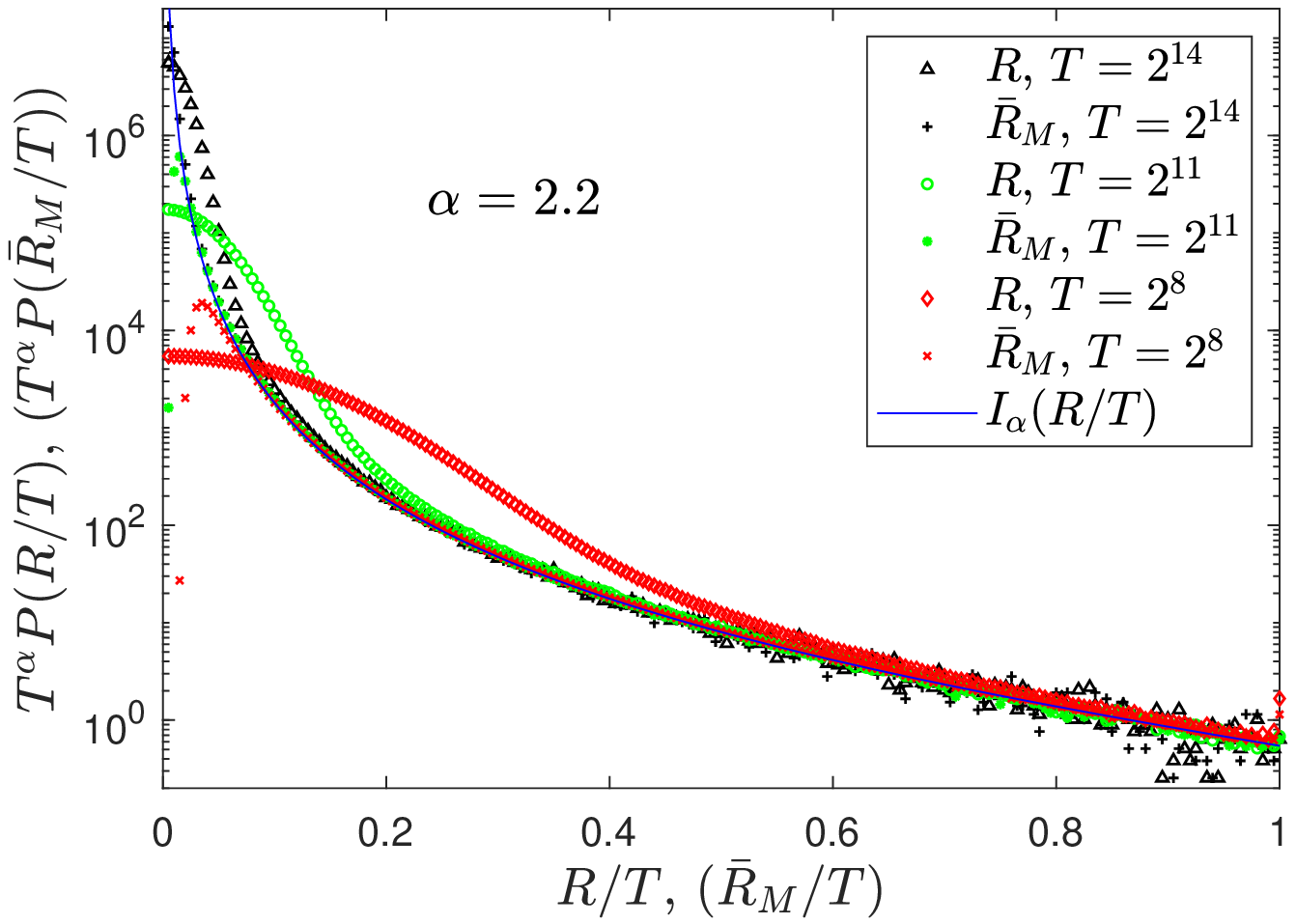}
	\caption{(Color on-line) Numerical verification of the big jump principle for the L\'evy Walk: $P(R,T)$ is compared to $P(\bar R_m,T)$ and to the analytic estimate 
		$I_{\alpha}(R/T)$. }.
	\label{LW2fig}
\end{figure}

\begin{figure}
	\includegraphics[width=0.45\textwidth]{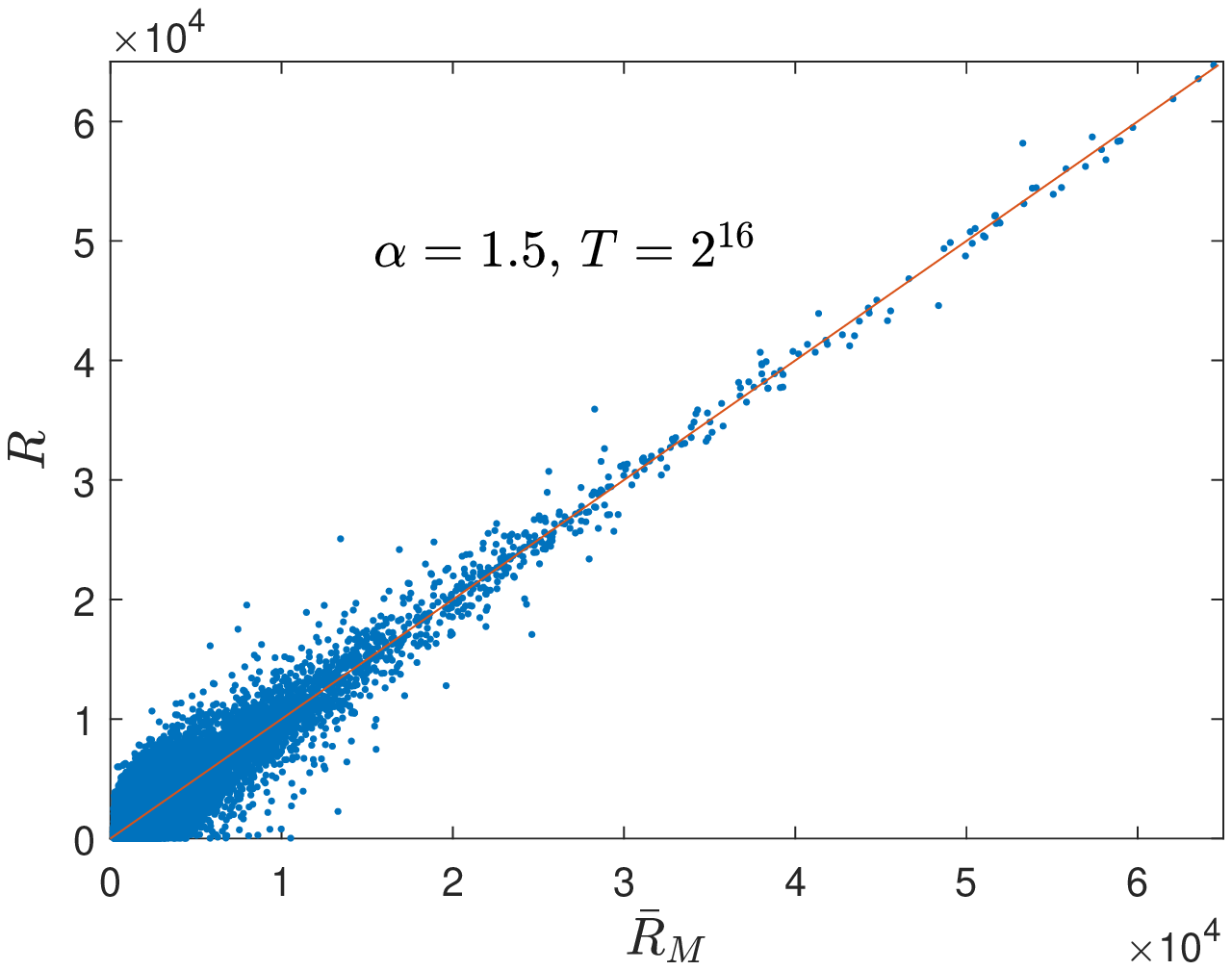}
	\includegraphics[width=0.45\textwidth]{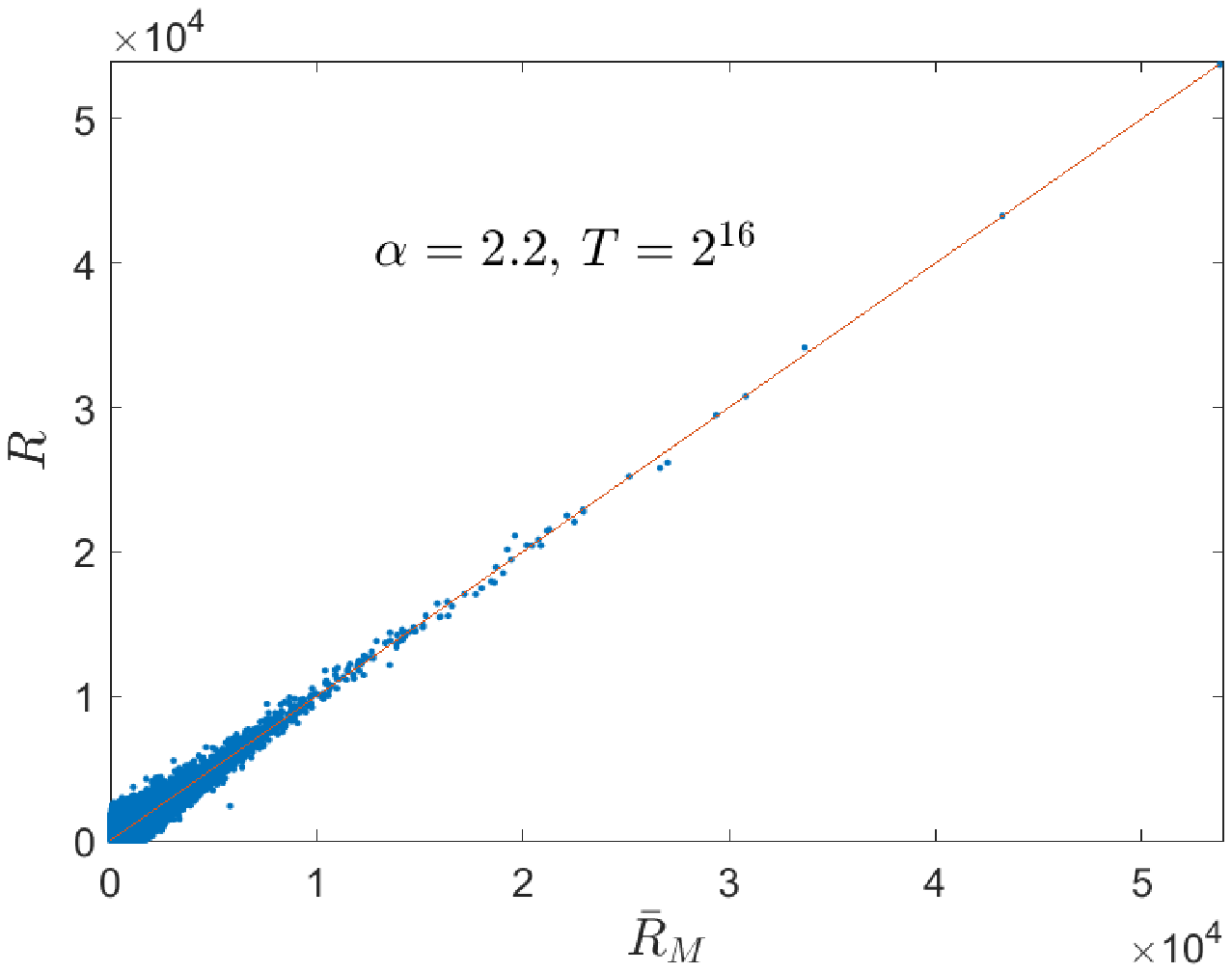}
	\caption{(Color on-line) Numerical verification of the direct correlation between the final position of the walker $R$ and  $\bar{R}_M=\max\{L_1,... L_B\}$. Dots represent the final position $R$ of the walker at $T=2^{16}$ as a function of the maximum jump  $\bar{R}_M$ of the same walker. Red line correspond to the linear plot $R=\bar{R}_M$. }
	\label{LW1fig}
\end{figure}

In Fig. \ref{LW2fig} we present our main results and compare them with finite
time simulations.  We plot the far tail of the PDF as a function of $R/T$, for $\alpha=1.5$, and $\alpha=2.2$. As expected, 
in the long time limit the densities  fully agree with the big jump approximation. 
We also plot the distribution of the largest jump $\bar{R}_M$. 
The agreement between
the distribution of $\bar{R}_M$ and the distribution of the total displacements
for large $R$ is clearly visible and both distributions
converge to the asymptotic results in a finite time scale. 
Fig. \ref{LW1fig} shows indeed that the biggest jump $\bar{R}_M$ and the final position $R$ are correlated: each dot represent $\bar{R}_M$ and $R$ for a single walker at $T=2^{16}$. For large $R$ we observe that $\bar{R}_M\approx R$, while for short distances large fluctuations are present due to multiple processes.

\section{Anomalous diffusion for cold atoms in optical lattices}
\label{atom}

\begin{figure}
\includegraphics[width=0.45\textwidth]{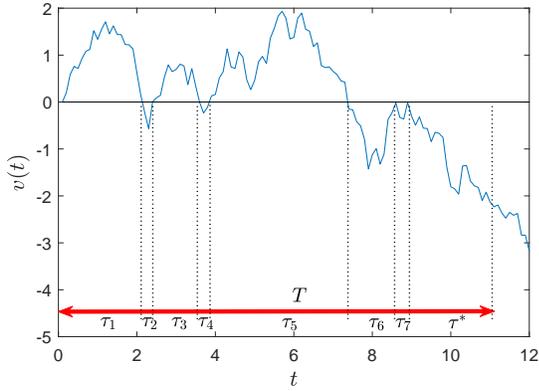}
\caption{(color online)
A realization of a path of the Langevin Eq. (\ref{atom1}) for an atom in an optical lattice.
The zero crossing define time intervals between renewal events since the underlying Langevin
process is Markovian. We show the waiting times, $\{\tau_1,....\tau^* \}$ which are known
to follow power law statistics, and in red we show the big jump. Notice that at observation time $T$ the process is not crossing zero, so the last time interval
called the backward recurrence time $\tau^*$, must be treated  differently.
For further details see:  \cite{Eli5}. 
}
\label{fig5}
\end{figure}

Both for  IID random variables and for L\'evy walks, the concept of "jump" is very clear: the displacement between renewals. But in real data we may have continuous trajectories, where the jump is not well defined (we are ignoring here sampling effects, which naturally lead to jumps. The topic of sampling is left for future work).  Hence, now we turn to a model known to generate L\'evy statistics, both in theory \cite{Eli3,Marksteiner} and in the lab \cite{Sagi}, based on a non-linear Langevin equation.  This is Sisyphus cooling \cite{Cohen,Castin,Marksteiner}.
Within this theory, energy dissipation of atoms in an optical traps can be described by the Langevin equation:

\begin{equation}
\dot v=-\frac{v}{1+v^2} + \sqrt{2D}\xi(t)
\label{atom1}
\end{equation}
where $\xi(t)$ is a white Gaussian noise with zero mean: $\langle\xi(t) \xi(t')\rangle=\delta(t-t')$ and $v$ represents the atom velocity.
In Eq. (\ref{atom1}) and all along this section we are using dimensionless variables
for velocity, time and space
(see details in \cite{Eli4}). The space $R$ covered by the atom in a time $T$ is:
\begin{equation}
R(T) = \int_0^T v(t) dt. 
\label{atom2}
\end{equation}
The motion of the atoms in this framework has been  studied in \cite{Eli3,Eli4,Eli5}. The dynamical evolution can be described in terms of a random walk where each step is defined by two subsequent events with $v(t)=0$, as described in Fig.  \ref{fig5}. Thus we will use the zero crossings of the velocity process to describe a jump and with this we will check the validity of the big jump principle \cite{note1}. More precisely the size of each jump is the area under the velocity curve between two zero crossings. 
Using this definition, according to equation (\ref{atom1}), the steps of the walker are uncorrelated but the duration and the length of each single step should be extracted by a complex distribution relating in a non trivial way space and time.  In particular, the joint distribution for a step having length $L$ and duration $\tau$ is
\begin{equation}
\Phi_E(L,\tau) = g(\tau)\phi_E(L|\tau),
\label{atom3}
\end{equation}
where $g(\tau)$ is the PDF for a step of duration $\tau$ and $\phi_E(L|\tau)$ is the conditional PDF for $L$ given $\tau$. In \cite{Eli4} it has been shown that for large $\tau$
\begin{equation}
g(\tau) \simeq g_* \tau^{-1-3\nu/2},
\label{atom4}
\end{equation}
where $g_*$ is a numerical constant and the exponent $\nu$ depends on the noise $D$ in eq. (\ref{atom1}) as:
\begin{equation}
\nu=\frac{1+D}{3D}. 
\label{atom5}
\end{equation}
The conditional PDF $\phi_E(L|\tau)$ obeys the following scaling property:
\begin{equation}
\phi_E(L|\tau)=\frac{1}{\tau^{3/2}}B_E(L/\tau^{3/2}).
\label{atom6}
\end{equation}
The scaling function $B_E(x)$ exponentially decays to zero at large $x$ and its analytic expression 
has been evaluated in \cite{Eli4}. Eq. (\ref{atom6})
shows that a step of length $L$ is covered in a time of order $\tau^{3/2}$. This accelerated 
motion corresponds to the ballistic motion of a single step in the L\'evy  walks. From Eqs. (\ref{atom1}-\ref{atom6}) we also get that the probability that a step has length $L$ is $q(L)\sim L^{-\nu-1}$.

In \cite{Eli3} the motion of the walker at short distances has been studied using techniques similar to the L\'evy walk case, with  $\nu$ playing the same role of the exponent $\alpha$. In particular for $\nu>2$ (i.e. $D<1/5$) both the mean square displacement $\langle L^2 \rangle$ and the mean duration $\langle \tau \rangle$ of a step are finite; therefore, $P(R,T)$ is a Gaussian with a characteristic length $\ell(T)\sim T^{1/2}$. For $2/3<\nu<2$ (i.e. $1/5<D<1$ ) the mean duration of a step is 
finite but the mean square displacement diverges; in this case $P(R,T)$ is described by a L\'evy scaling function $L_\nu(\cdot)$ i.e. $P(R,T)=\ell(T)^{-1}L_\nu(R/\ell(T))$; where $\ell(T)\sim T^{1/\nu}$. Finally for $\nu<2/3$ (i.e. $D>1$) the motion is accelerated as $T^{3/2}$ and $P(R,T)=\ell(T)^{-1} f_{\nu}(R/\ell(T))$ with $\ell(T)\sim T^{3/2}$ ($f_{\nu}(\cdot)$ is a $\nu$ dependent scaling function). 

In the case $\nu>2/3$ we expect that the probability of finding a particle in a position $R\sim T^{3/2} \gg \ell(T) $ can be evaluated considering a single big jump leading it to a distance  $L\sim T^{3/2}\gg \ell(T)$. In particular we can consider, at time $T_w$, the probability $p_{{\rm tot}}(L,T_w,\tau)dLdT_wd\tau$ that the particle makes a jump of length $L$ and duration $\tau$: taking into account that the probability of making a step is $\langle \tau \rangle^{-1}$ independently of $T_w$. We have:
\begin{equation}
p_{{\rm tot}}(L,T_w,\tau)=\frac{\Phi_E(L,\tau)}{\langle \tau \rangle}. 
\label{atom7}
\end{equation}
The PDF $P(R,T)$ can be calculated taking into account the different processes driving the particle at a distance $R$ at time $T$ with a single jump of length $L$. As in the case of  L\'evy walks there are two possibilities. First the particle can make a jump of length $L=R$ and duration $\tau$; such a jump can be made at any $T_w\in[0,T-\tau]$. Moreover all the values of $\tau\in[0,T]$ have to be taken into account. Since $dR=dL$, we get the contribution of this process to $P(R,T)$ by integrating over all possible $\tau$ and $T_w$:
\begin{eqnarray}
\int_0^T d\tau \Phi_E(R,\tau) \int_{0}^{T-\tau} \frac{dT_w}{\langle \tau \rangle}
&\sim&  \nonumber\\
\int_0^T d\tau \frac{g_* (T-\tau)}{\langle \tau \rangle \tau^{\frac{5}{2} + \frac{3 \nu}{2}}}B_E\left(\frac{R}{\tau^{\frac{3}{2}}}\right)&=&B_1(R,T).
\label{atom8}
\end{eqnarray}
where the second expression holds for large $R$ and $T$.
In this case, the single step is characterized by a superballistic motion where in a time $T$ the particle covers a distance of order $T^{3/2}$, therefore the natural rescaled variable is $z=R/T^{3/2}$. Defining  
$y=R/\tau^{3/2}$ we get:
\begin{equation}
B_1(z,T)= \frac{g_*}{T^{\frac{1}{2} + \frac{3 \nu}{2}}\langle \tau \rangle z^{1+\nu}}\int_z^\infty dy  \left(1-\left({z}/{y}\right)^{\frac{2}{3}} \right)
y^\nu B_E\left(y \right) 
\label{atom9}
\end{equation}

As for the L\'evy walk, another kind of processes provides a non trivial contribution to $P(R,T)$, i.e. when at time $T$ the walker is still moving in the big jump.
In this case, one has to consider the probability to perform a jump longer than $R$. Since the motion is the result of a Langevin stochastic process, the distance $R$ can be covered in different times $\tau^*$, hence,
we call $\Psi_M(R,\tau^*)$ the probability  to cover in a step  a distance larger than $R$ arriving in $R$ exactly at $\tau^*$. According to \cite{Eli3,Eli5} we can write:
\begin{equation}
\Psi_M(R,\tau^*)=w(\tau^*)\frac{1}{\tau^{*\frac{3}{2}}}B_M\left(\frac{R}{\tau^{*\frac{3}{2}}} \right)  
\label{atom10}
\end{equation}
where $w(\tau^*)=\int_{\tau^*}^{\infty} g(\tau)d\tau$ is the probability of making a jump of duration longer than $\tau^*$, while $\psi_M(R|\tau^*)=\tau^{*-{3}/{2}} B_M({R}/{\tau^{*{3}/{2}}} )$ is the conditional probability of covering a distance larger than $R$ given $\tau^*$. We remark that $\Psi_M(R,\tau^*)$ can be introduced also in L\'evy walks where ballistic motion entails that trivially: $\Psi_M(R,\tau^*)=\delta(R-v\tau^*)\int_{R}^{\infty} \lambda(L) dL$. On the other hand, if the motion during the step is determined by Eq. (\ref{atom1}), $B_M(\cdot)$ displays a non-trivial behavior (see \cite{Eli4} for details). 
We notice that only the jumps occurring at $T_w=T-\tau^*$ bring the walker in $R$ at time $T$. Therefore, integration over $T_w$ is not necessary or equivalently we can insert a delta function. However, different $\tau^*$ provide different contributions to the process, so we have to integrate over the possible $\tau^*\in [0,T]$. Taking into account that the jumping rate $\langle \tau \rangle^{-1}$ is independent of $T_w$, the contribution to the PDF is:
\begin{eqnarray}
\int_{0}^{T} 
{d\tau^*} \Psi_M(R,\tau^*) \int_{0}^{T} dT_w \frac{\delta(T_w-T+\tau_*)}{\langle \tau \rangle} 
&\sim&  \\ \frac{g_*}{\langle \tau \rangle}\frac{2 }{3 \nu}
\int_{0}^{T} d \tau^* \frac{1}{\tau^{*\frac{3}{2}(1+\nu)}}B_M\left(\frac{R}{\tau^{*\frac{3}{2}}} \right)
&=&B_0(R,T) \nonumber
\label{atom11}
\end{eqnarray}
where we take into account that for large $\tau^*$ we have 
$w(\tau^*)\sim g_* \int_{\tau^*}^{\infty} \tau^{-1-3\nu/2} d\tau=g_* (2/3\nu)\tau^{*-3\nu/2}$.
Introducing in (\ref{atom11}) the rescaled variable $z=R/T^{3/2}$ and defining  
$y=R/\tau^{*3/2}$ we get:
\begin{equation}
B_0(z,T)= \frac{1}{T^{\frac{1}{2} + \frac{3 \nu}{2}}}
\frac{2 g_*}{ 3 \nu \langle \tau \rangle z^{1+\nu}} 
\int_z^\infty dy  \left({z}/{y}\right)^{\frac{2}{3}} y^\nu B_M\left(y \right) 
\label{atom12}
\end{equation}
Summing the contributions to $P(R,T)$ in Eqs (\ref{atom9},\ref{atom12}) we get 
\begin{equation}
P(R,T)\sim B_0(z,T)+B_1(z,T)= \frac{1}{T^{\frac{1}{2} + \frac{3 \nu}{2}}}
I_\nu(R/T^{\frac{3}{2}})
\label{atom13}
\end{equation}
i.e. the expression obtained in \cite{Eli5} with a totally different method. In 
\cite{Eli5} a comparison of Eq. (\ref{atom13}) with numerical simulations 
is also presented showing a very nice agreement in the asymptotic regime.
As discussed in details in \cite{Eli5}, Eq. (\ref{atom13})  is not normalized, since $I_\nu(x)$ diverges for $x \to 0$. This again is hardly surprising since Eq. (\ref{atom13}) works for large values of $R$. Still the big jump principle provides  the moments of order $q>\nu$,  and as such  
it gives the   infinite density of the process (like the simpler L\'evy walk case). We remark that also in this case the long tails are described by the same scaling length characterizing the single jump dynamics i.e. $T^{3/2}$ which plays the same role of the ballistic motion in the L\'evy walk case.

\section{The Single big jump in the L\'evy-Lorentz gas}
\subsection{The L\'evy Lorentz gas}

The approach introduced for the  L\'evy walk, that takes into account the different contributions to the big jumps in the PDF, can be applied to the case of a walker moving in a random sequence of $1$-D scatterers spaced according to a L\'evy distribution \cite{klafter}, i.e. a L\'evy-Lorentz gas. This is a prototypical model where  the highly non trivial correlation among the steps is introduced by the quenched disorder, that is the positions of the scatterers in a sample.
We build the system placing a scatterer on the origin and spacing the others in the positive and negative directions so that the  probability density for two consecutive scatterers to be at distance $L$ is $\lambda(L)$ as defined in Eq. (\ref{lambdaL}). A continuous time random walk \cite{ctrw} 
is naturally defined on the $1$-D quenched scatterers distribution: 
a walker starts from the scatterer in the origin, then it moves with constant velocity $v$ until it reaches one of 
the scatterers, and then it is transmitted or reflected with probability $1/2$.
We consider walkers starting from a scattering point. Indeed it is known that for $\alpha<1$,
i.e., when the average distance between scatterers diverges,  the results in the asymptotic region depend on the initial position of the walker \cite{klafter,levycant}.
Moreover, the PDF to be at distance $R$ from the origin at time $T$ $P(R,T)$ has been obtained by averaging both on the walker trajectories  and on the realizations of the disorder.

\begin{figure*}
	\includegraphics[width=0.45\textwidth]{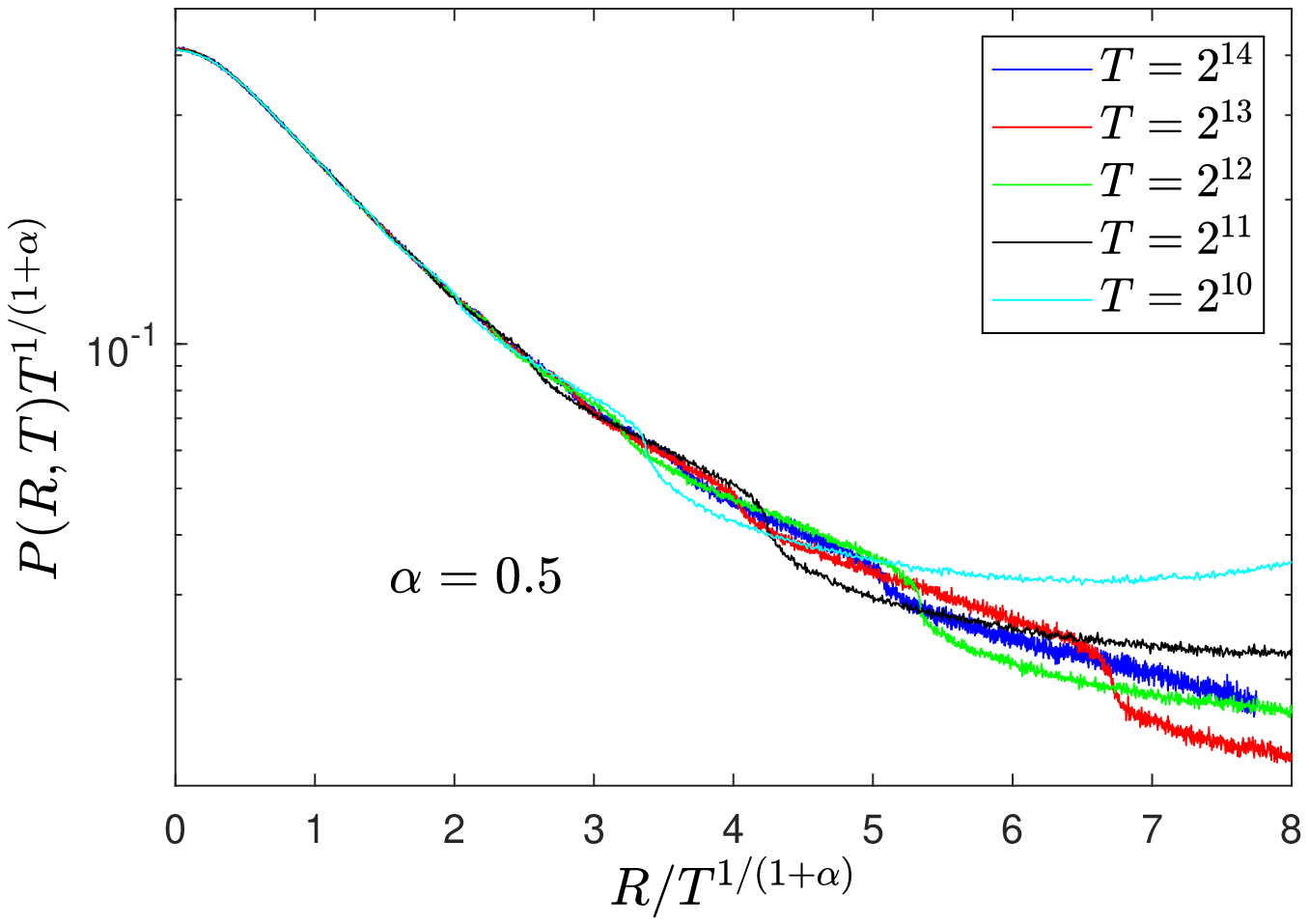}
	\includegraphics[width=0.45\textwidth]{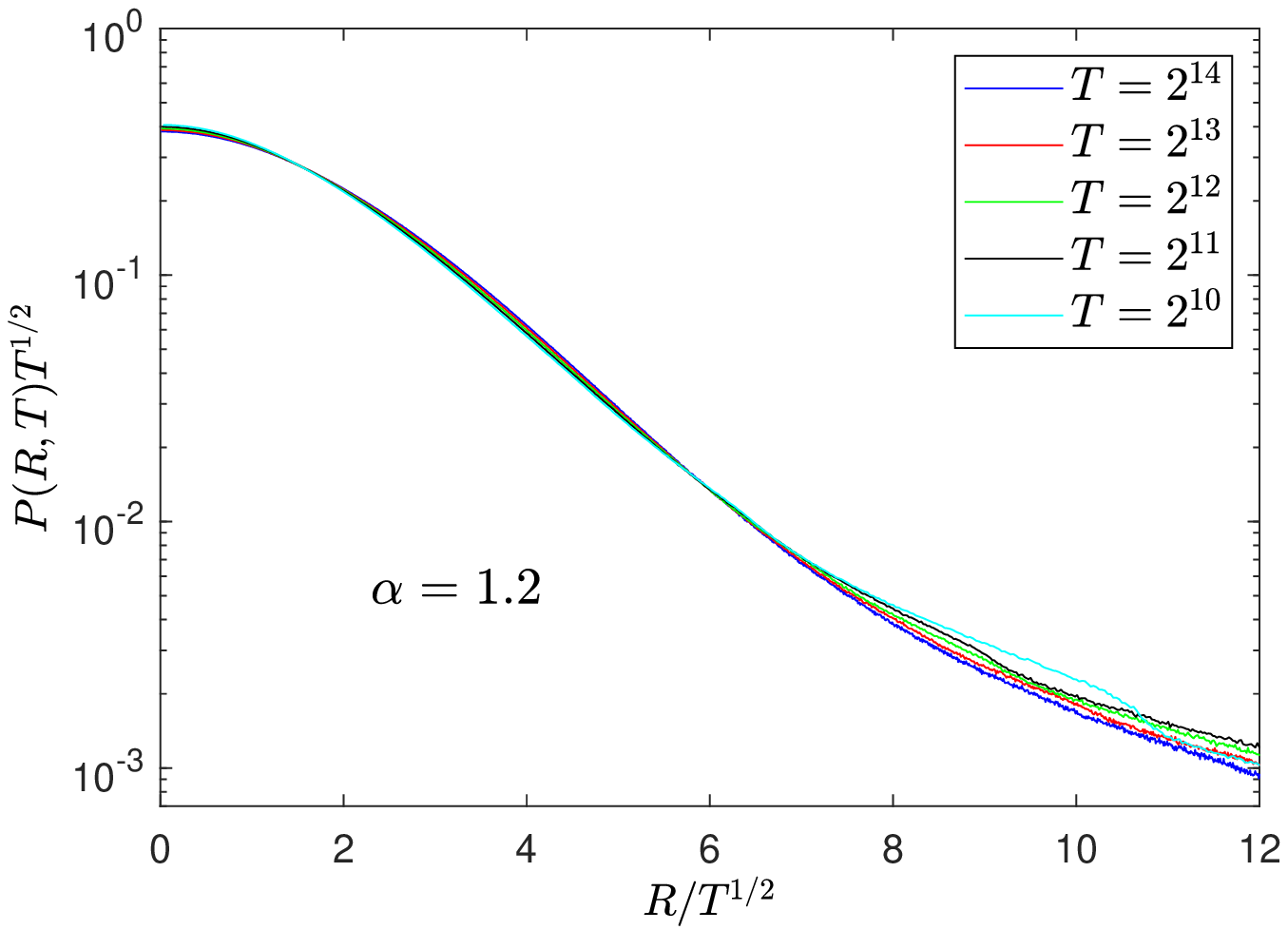}
	\includegraphics[width=0.45\textwidth]{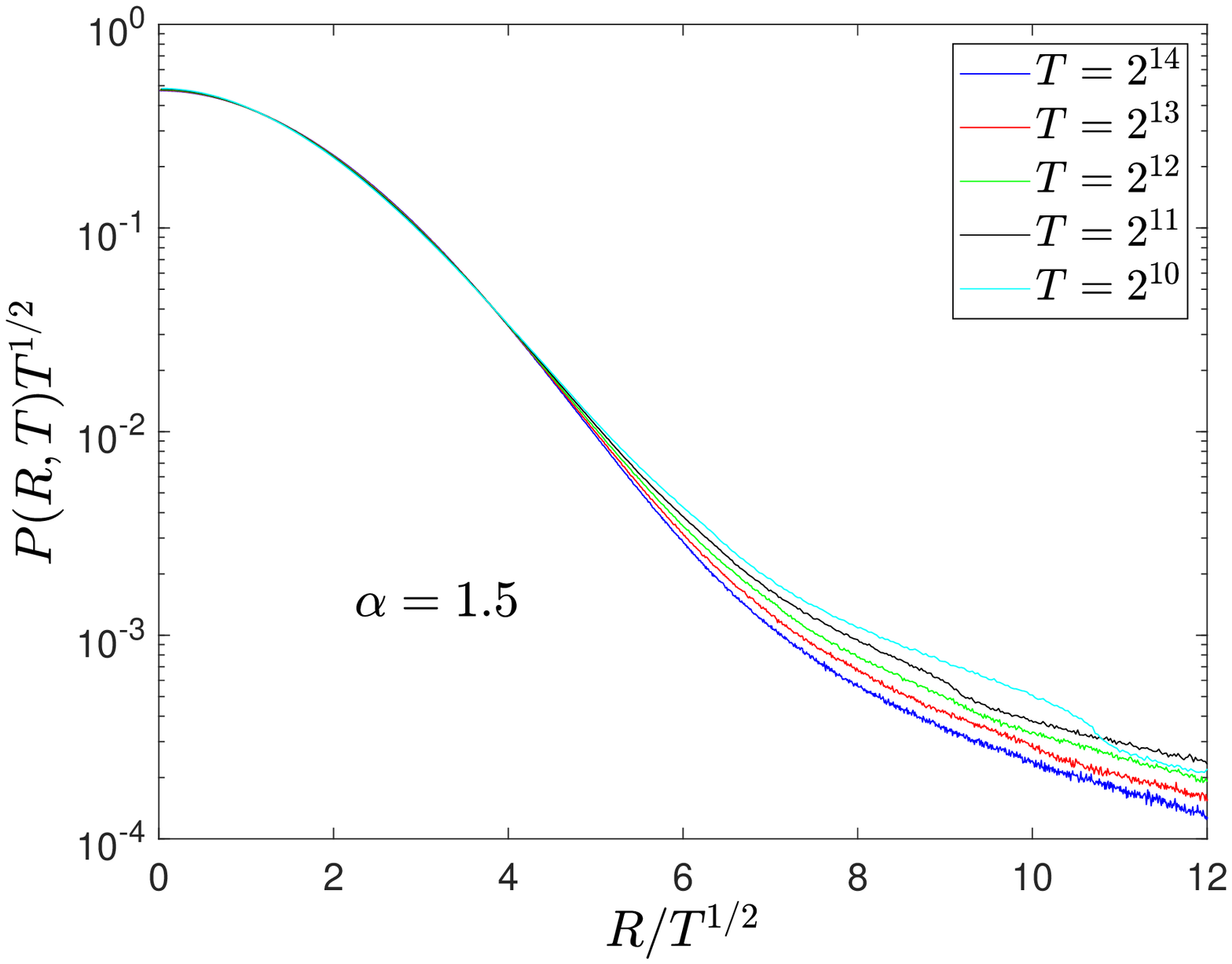}
	\includegraphics[width=0.45\textwidth]{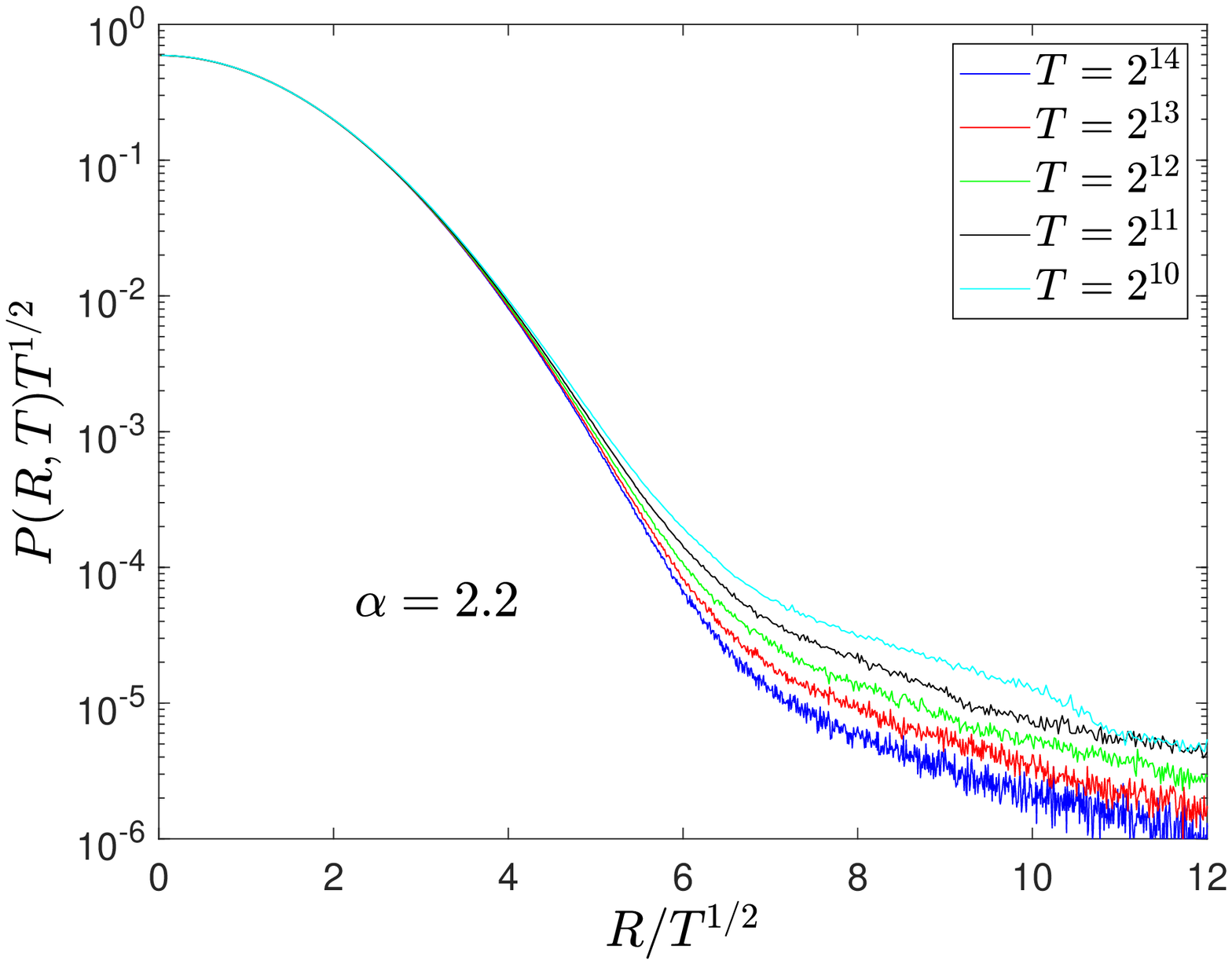}
	\caption{(Color on-line) Short distance scaling of the PDF for the  L\'evy-Lorentz gas according to Eqs. (\ref{ellcom},\ref{sal}). A Gaussian PDF is observed for $\alpha>1$. }.
	\label{scalSH}
\end{figure*}

In \cite{levyrand,levysanto}, using an analogy with an equivalent electrical model 
\cite{beenakker}, it has been shown that the bulk part of $P(R,T)$ 
displays a scaling behavior with a characteristic length $\ell (T)$ growing as:
\begin{equation}
\ell (T) \sim
\begin{dcases}
T^{\frac{1}{1+\alpha}} & \mathrm{if}\ 0<\alpha<1 \\
T^{1 \over 2} & \mathrm{if}\ \alpha>1  
\label{ellcom}
\end{dcases}
\end{equation}
In particular the scaling form of $P(R,T)$ reads:
\begin{equation}
P(R,T)=\ell^{-1}(T)f(R/\ell(T))+B(R,T) 
\label{sal}
\end{equation}
with a convergence in probability to $\ell^{-1}(T)f(R/\ell(T))$
\begin{equation}
\lim_{T\to\infty}\int_0^{\infty} |P(R,T)-\ell^{-1}(T)f(R/\ell(T))|dR=0.
\label{sal2}
\end{equation}
The leading contribution to $P(R,T)$  is hence  
$\ell^{-1}(T)f(R/\ell(T))$, which is significantly different from 
zero only for $R \lesssim \ell(T)$. The short distance behavior described by Eqs. (\ref{ellcom}-\ref{sal2}) has been tested in numerical simulations,
as shown in Fig. \ref{scalSH}, and then rigorously proved in a series of  recent papers \cite{Bianchi,Magdziarza,Bianchi2,Artuso}.

The subleading term $B(R,T)$ (that satisfies  $\lim_{T\to \infty}\int | B(R,T)| dR=0$), describes the behavior of $P(R,T)$
at larger distances, i.e. $\ell(T)\ll R < v T$ (since the velocity $v$ is finite,  $B(R,T)$ is 
strictly zero for  $R>vT$). Notice that $B(R,T)$ can provide important contributions to higher moments of the distribution:
\begin{eqnarray}
\langle R^q (T)\rangle & = &  \int_0^{K \ell(T)} \ell^{-1}(T)f(R/\ell(T)) R^q dR + \nonumber \\ & \ &+ \int_{K \ell(T)}^{vT} B(R,T) R^q dR.
\label{rP}
\end{eqnarray}
where $K$ is a finite constant, as the first term can be sub-leading with respect the second integral for large enough $q$.  Notice that Eq. (\ref{rP}) contains once again the natural cut off  $vT$, that is the maximum distance that the walker can cover in a time $T$.  This suggests that the ballistic scaling length $vT$ characterizing each single step becomes dominant at large distance. 

We will show that $B(R,T)$ exhibits the following scaling:
\begin{equation}
B(R,T) \sim
\begin{dcases}
T^{-{1+\alpha+\alpha^2 \over 1+\alpha}}  I_{\alpha}\left({R \over vT} \right) & \mathrm{if}\ 0<\alpha<1 \\
T^{-{1 \over 2} -\alpha} I_{\alpha}\left({R \over vT} \right) & \mathrm{if}\ 1 \leq \alpha  
\label{single_j2}
\end{dcases}
\end{equation}
where $I_{\alpha}(x)$ is an $\alpha$-dependent scaling function that we will evaluate analytically using the big jump principle. The single jump dynamics gives rise to a ballistic scaling length  $vT$.  We will show that $\int I_{\alpha}(x) dx =\infty$ and therefore $I_{\alpha}(.)$ is an infinite density \cite{Eli1,Eli2} and as discussed in Eq. (\ref{rP}) for large enough $p>0$, $I_{\alpha}(x)$ can be used to estimate the moments of the process.
In particular, the competition of time scales in Eqs. (\ref{ellcom}-\ref{single_j2}) provides the full behavior of $\langle R^q(T) \rangle$ as a function of time \cite{levyrand,Giberti}:
\begin{equation}
\langle R^q (t) \rangle \sim
\begin{cases}
T^{\frac{q}{1+\alpha}}\sim \ell(T)^q & \mathrm{if}\ \alpha<1,
\ q<\alpha \\
T^{\frac{q(1+\alpha)-\alpha^2}{1+\alpha}} & \mathrm{if}\ \alpha<1, 
\ q>\alpha \\
T^{\frac{q}{2}}\sim \ell(T)^q & \mathrm{if}\ \alpha>1,
\ q<2 \alpha-1 \\
T^{\frac{1}{2}+q-\alpha} & \mathrm{if}\  \alpha>1,
\ q>2 \alpha-1 
\label{rpc2}
\end{cases}.
\end{equation}
In \cite{levyrand} the asymptotic behaviors of Eq.s (\ref{rpc2}) has been obtained using a single big jump heuristic argument
and it has been shown that the results are consistent with numerical simulations. Similar estimates, always based on single big jump arguments,
have been obtained in higher dimensions \cite{buons,Ub}.

\begin{figure*}
	\includegraphics[width=0.45\textwidth]{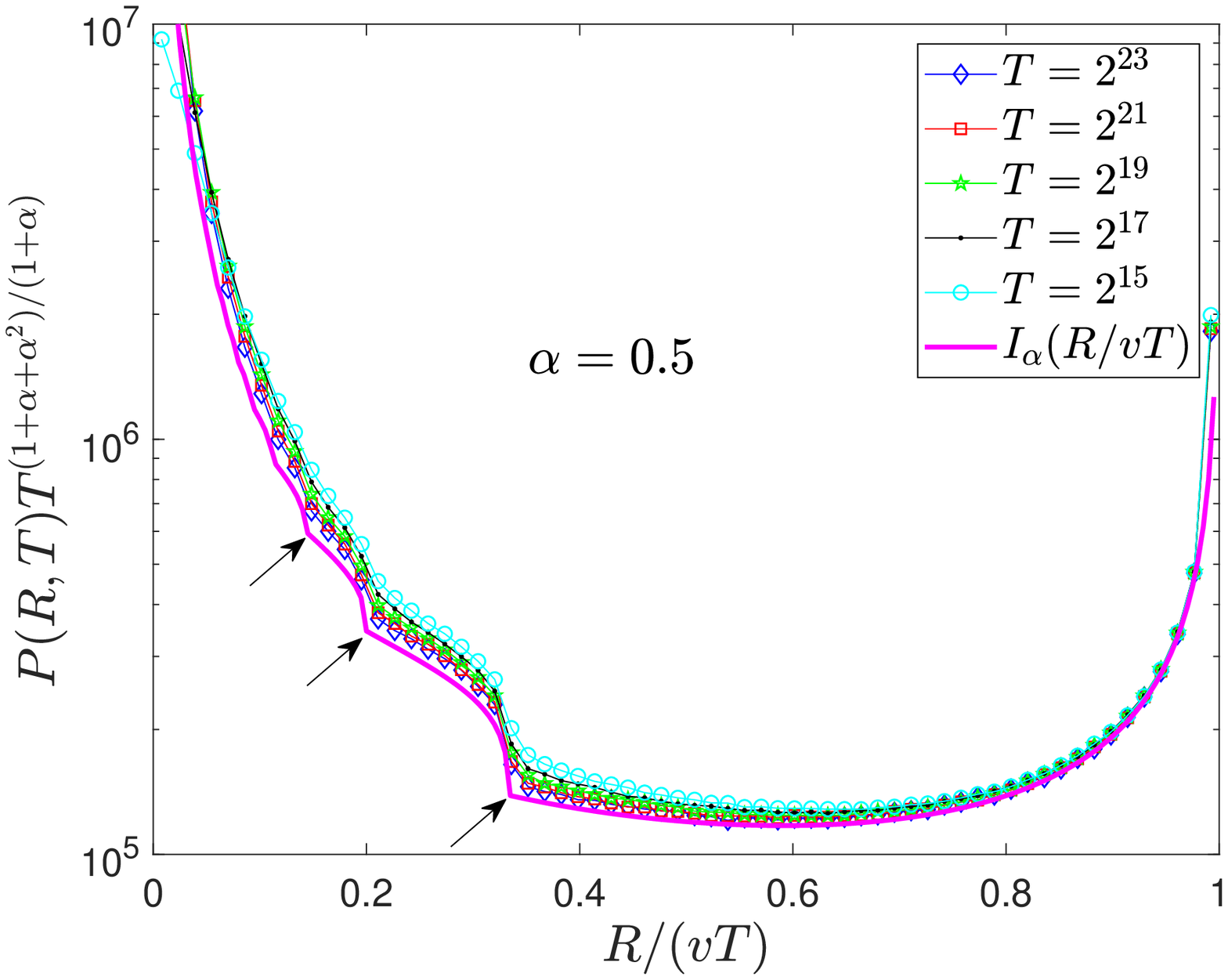}
	\includegraphics[width=0.45\textwidth]{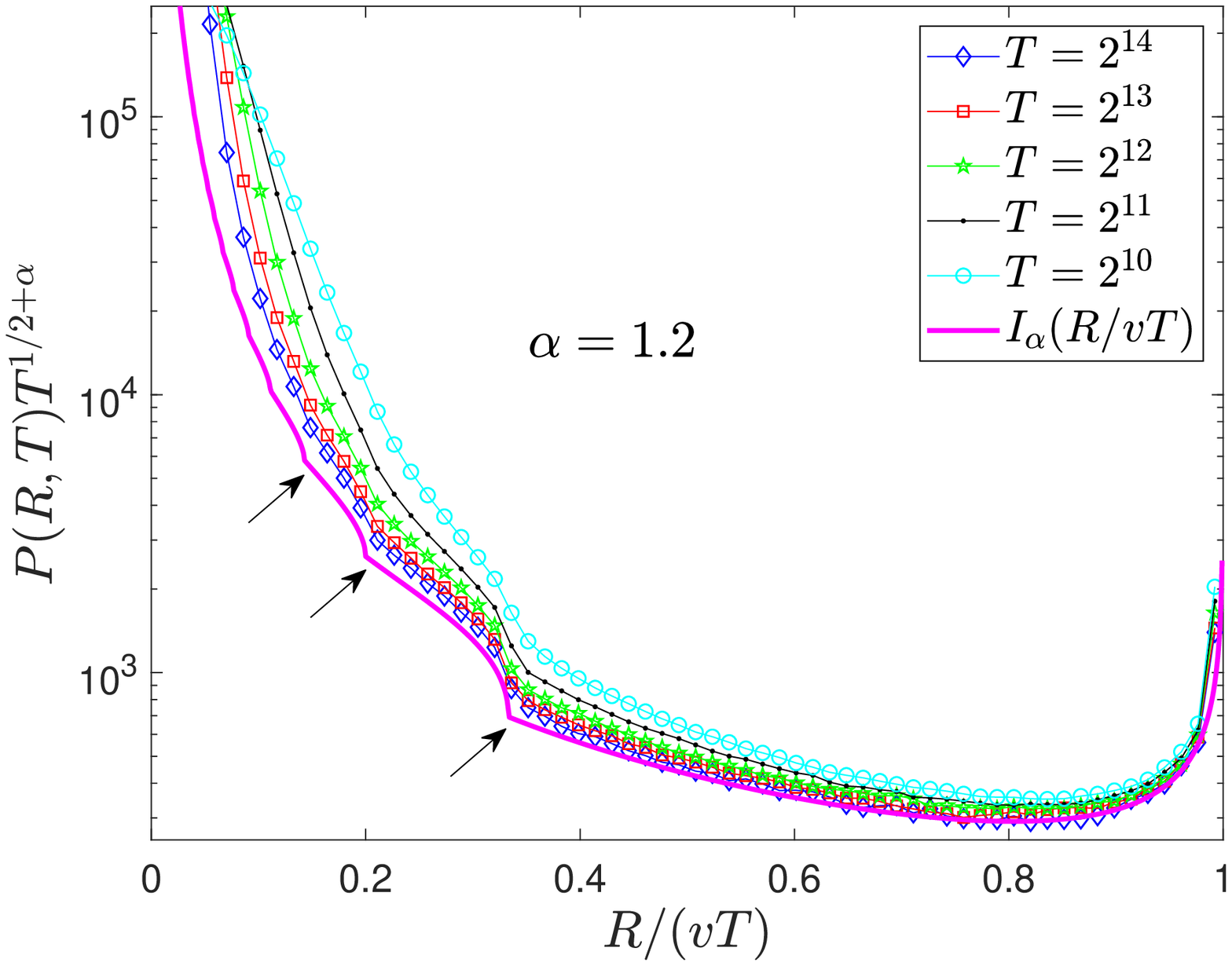}
	\includegraphics[width=0.45\textwidth]{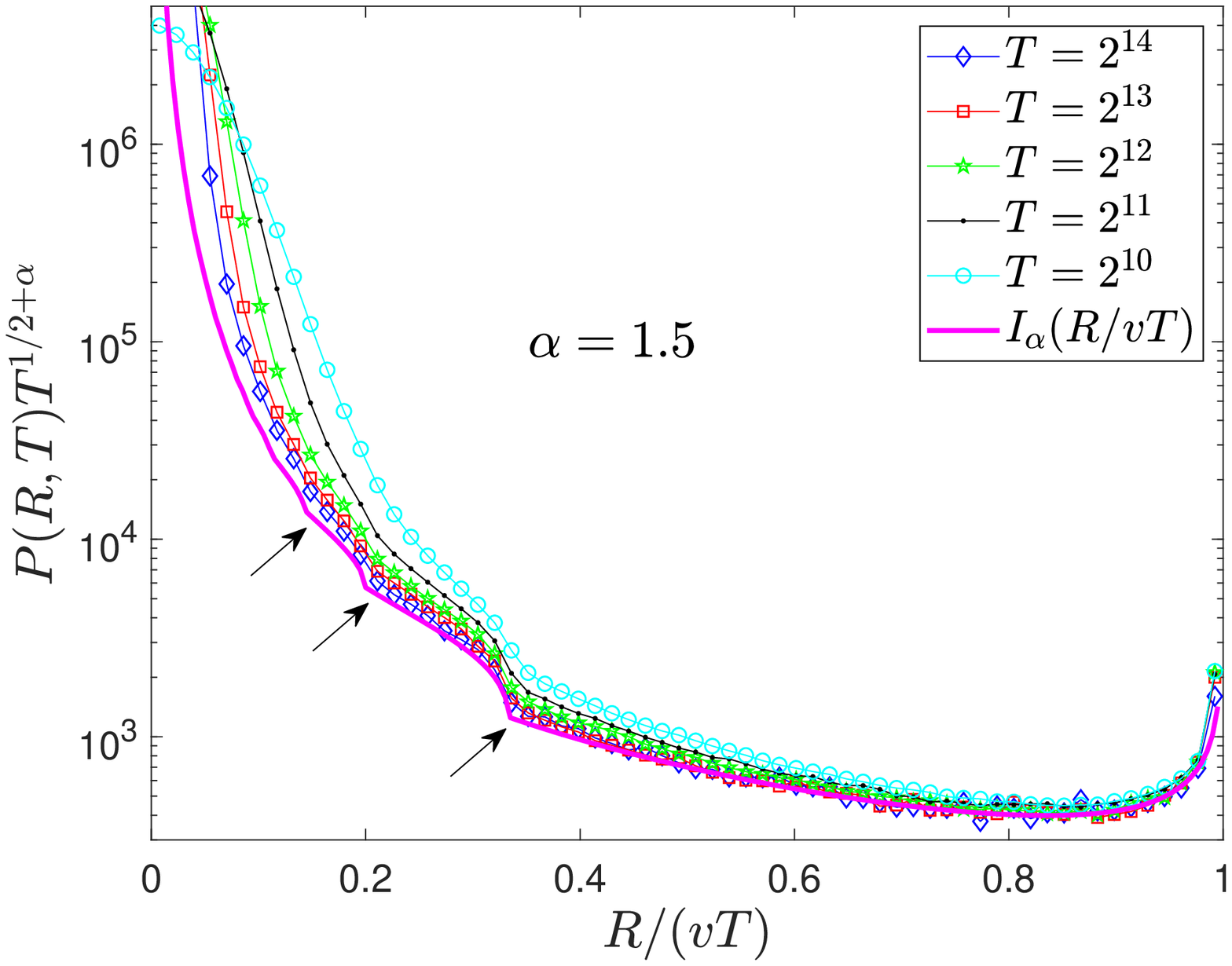}
	\includegraphics[width=0.45\textwidth]{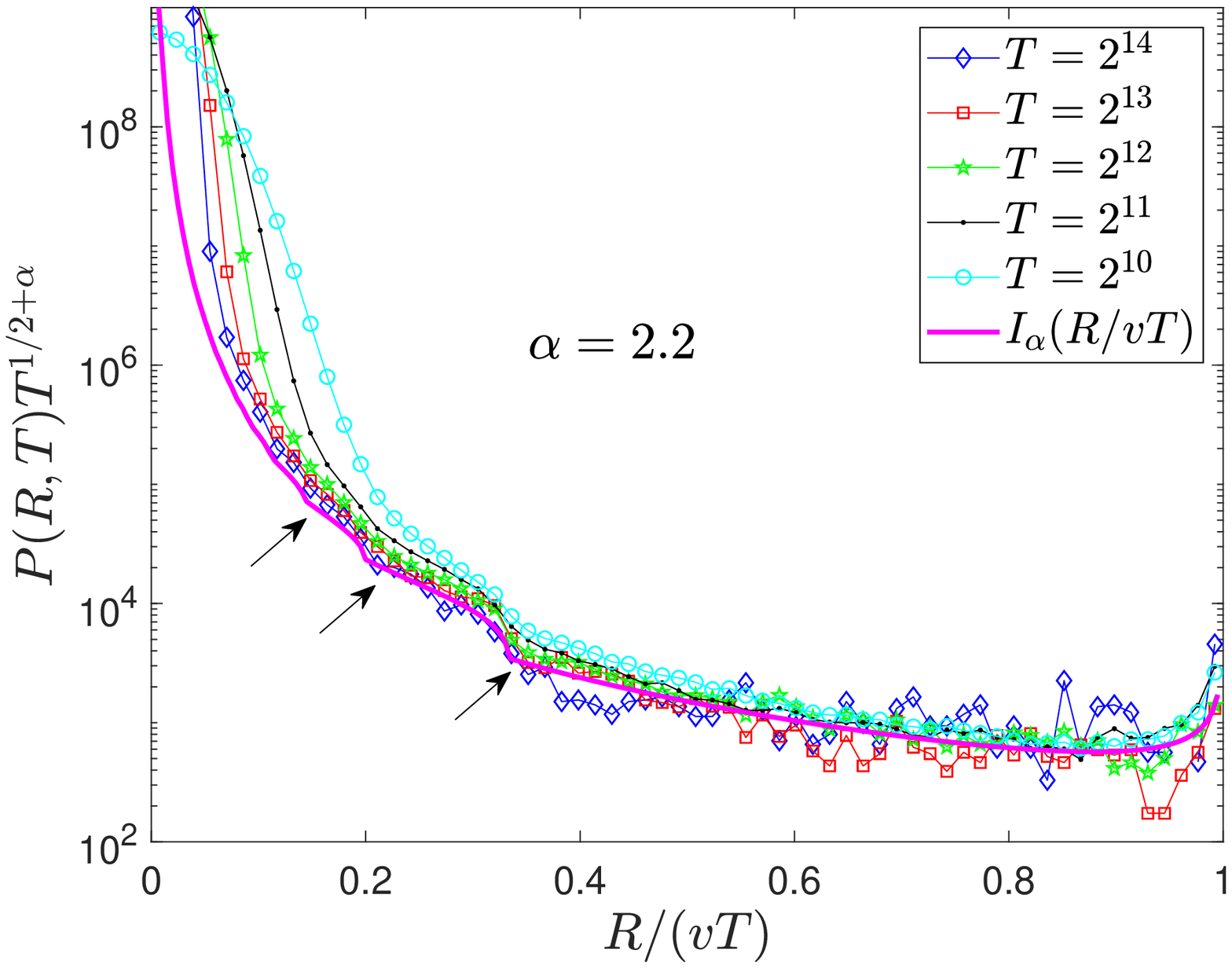}
	\caption{(Color on-line) Rescaling of the PDF for the L\'evy-Lorentz gas according to Eq. (\ref{gtot}). The data at $\alpha=0.5$, $\alpha=1.2$ $\alpha=1.5$ and $\alpha=2.2$ are obtained by averaging over $5\cdot 10^7$, $5\cdot 10^8$, $10^9$ and $4\cdot 10^9$ walkers moving on different realizations of the structure. Notice the discontinuities at $R/(vT)=1/3,1/5,1/7 ...$ which are found  for all values of $\alpha$. }.
	\label{scalp}
\end{figure*}

\begin{figure*}
	\includegraphics[width=0.45\textwidth]{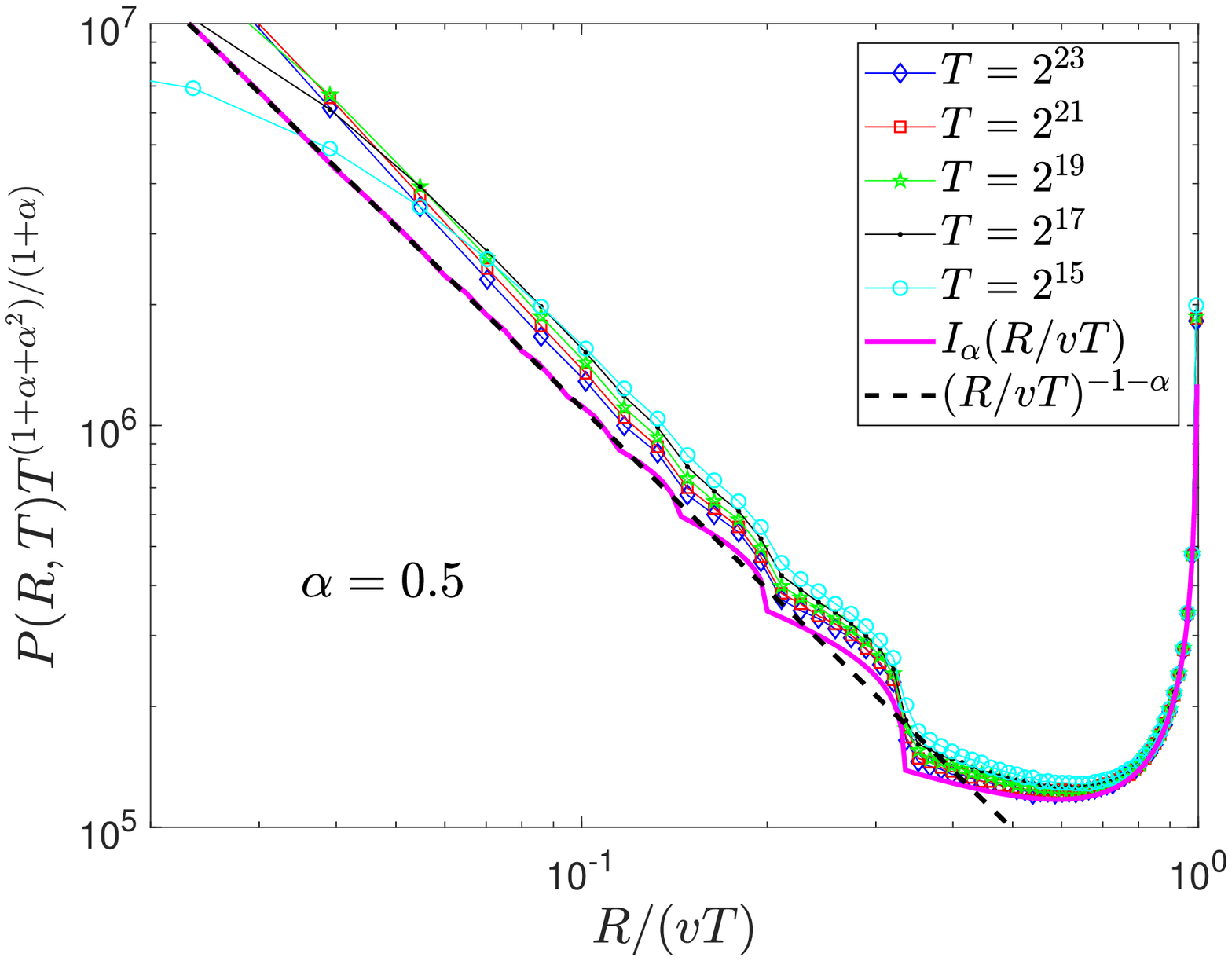}
	\includegraphics[width=0.45\textwidth]{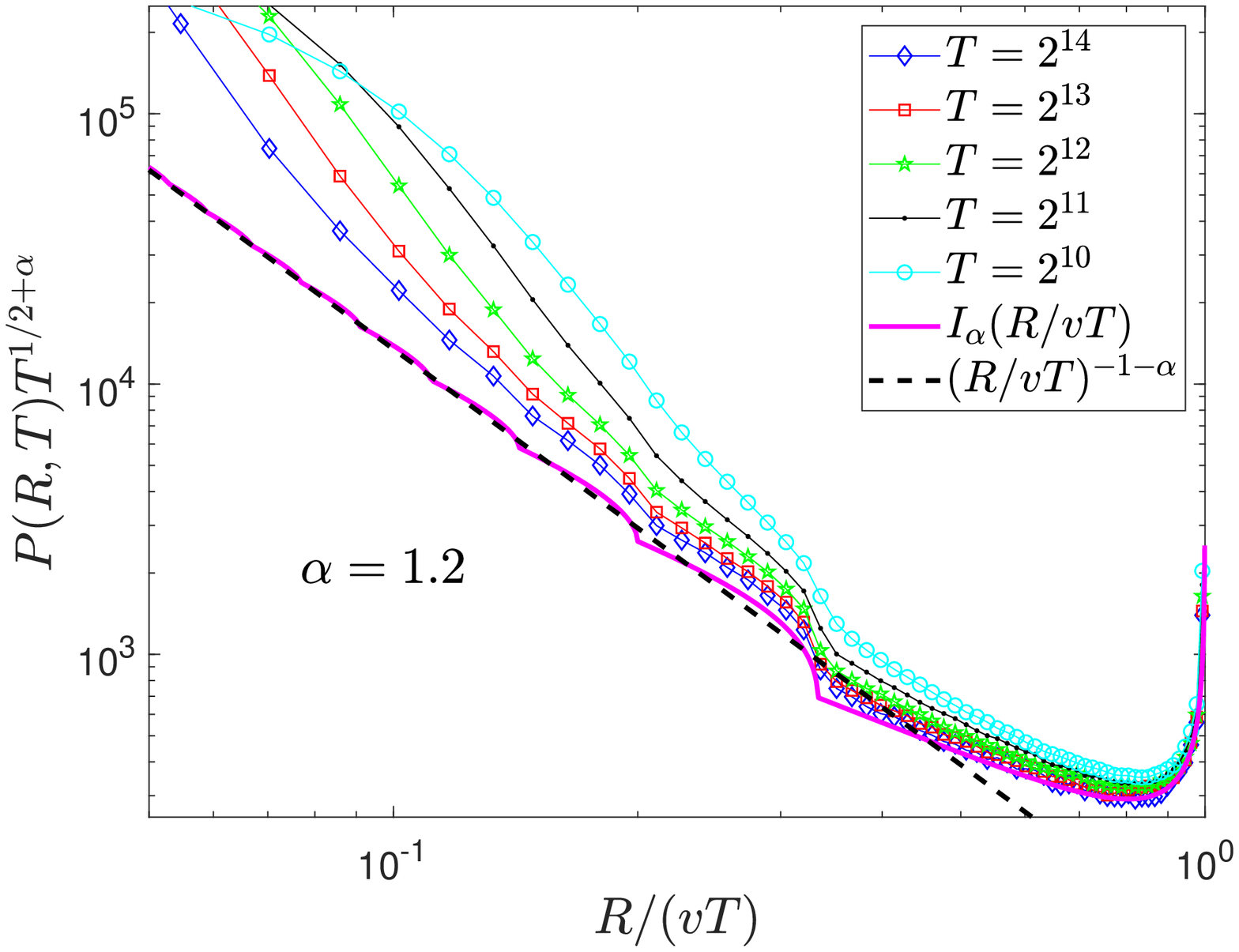}
	\includegraphics[width=0.45\textwidth]{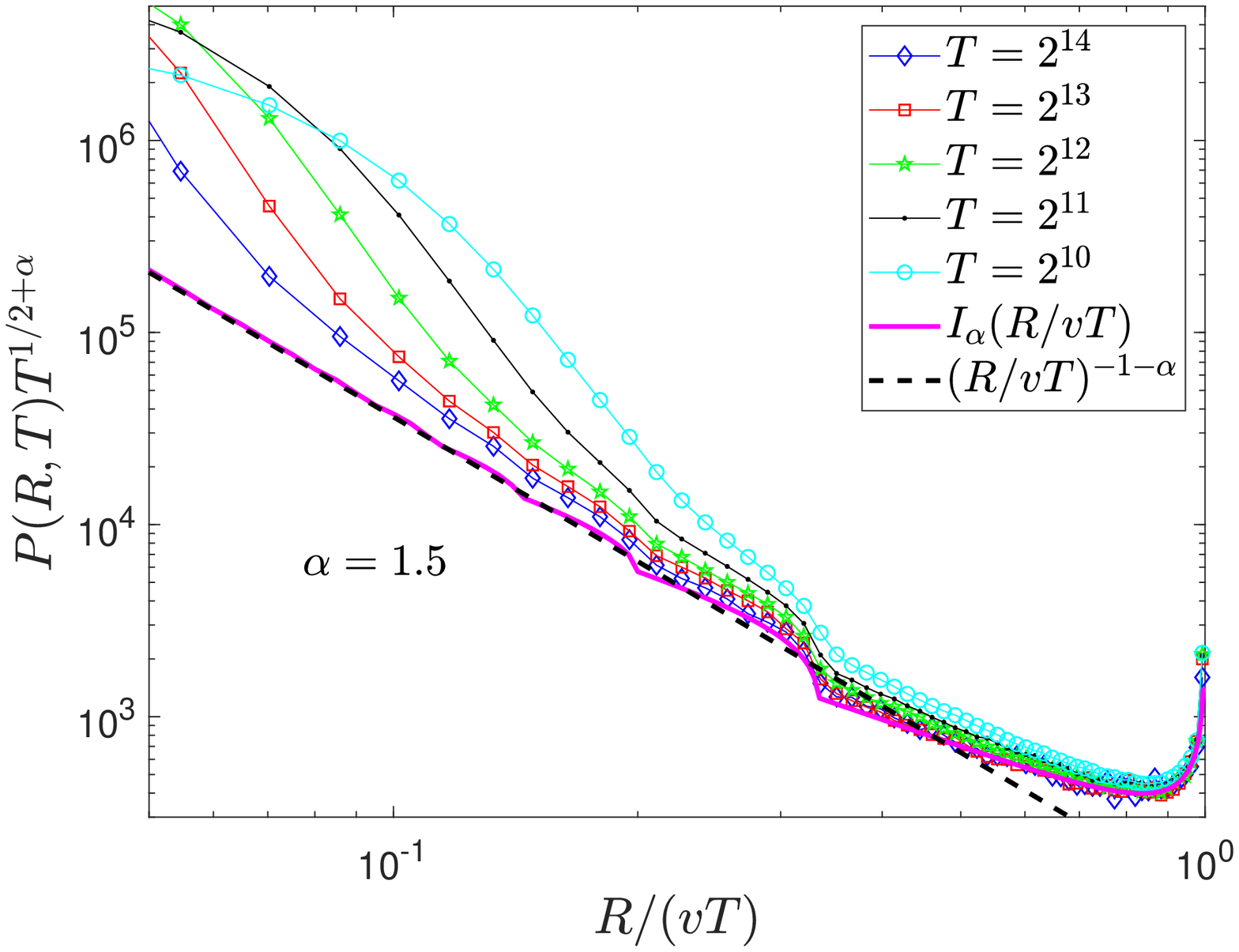}
	\includegraphics[width=0.45\textwidth]{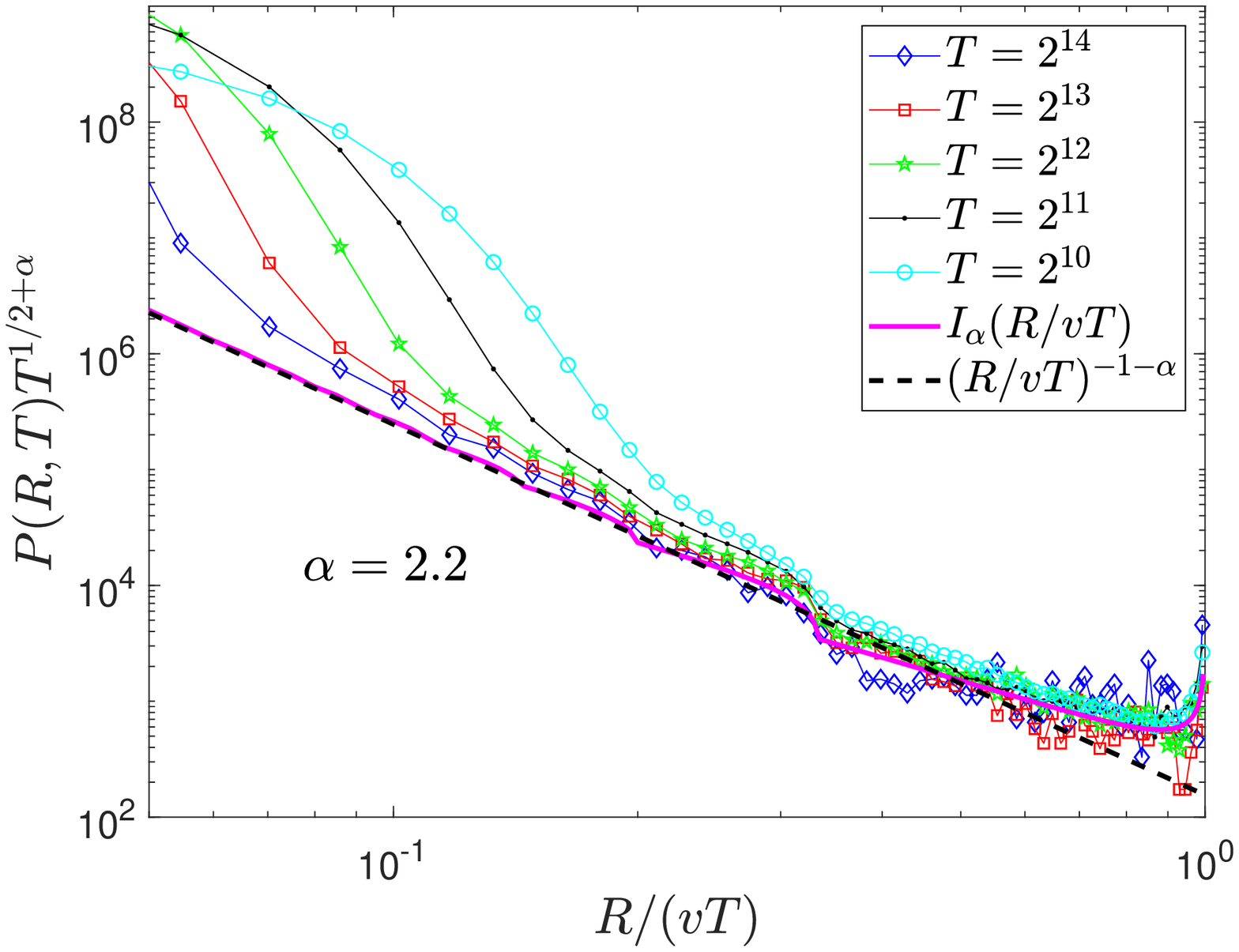}
	\caption{(Color on-line) The same data of Fig. \ref{scalp} plotted in a log-log scale.
		Here the approach of the scaled solution to the non-normalizable  $R^{-1-\alpha}$ pole on the origin is visible. }
	\label{scalpLog}
\end{figure*}

However, $I_{\alpha}(x)$ is not merely a mean to generate moments, as it describes the far tails of $P(R,T)$.  In particular, numerical simulations presented in Fig. \ref{scalp}  show that Eq. (\ref{single_j2}) provides the correct scaling behavior for $P(R,T)$ at large $R$. Furthermore it is evident from these figures that the far tail of the spreading particles is non-trivial in the sense that the packet exhibits non analytical behaviors and surprising step like structures.

\subsection{An analytical estimate of the big jump}

\begin{figure}
	\includegraphics[width=0.45\textwidth]{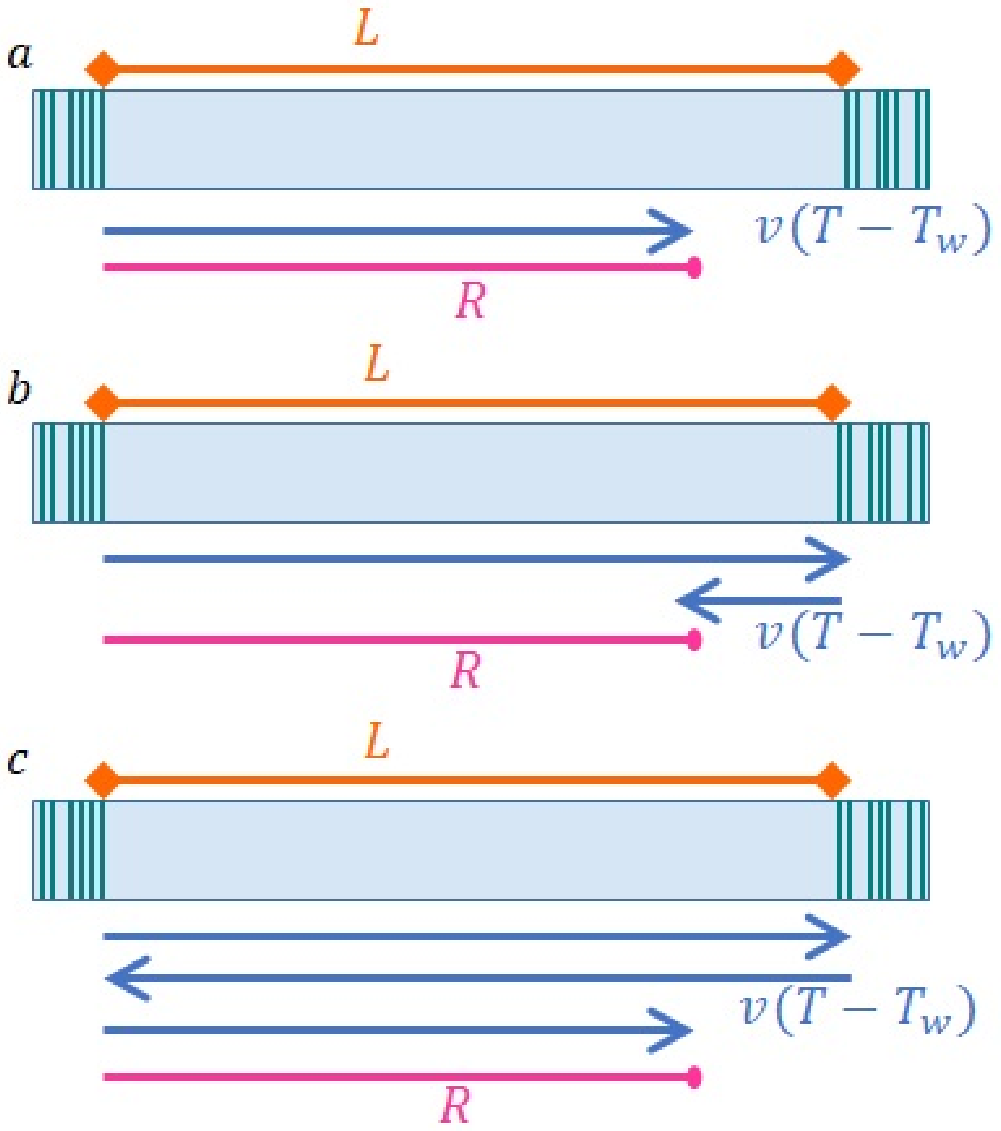}
	\caption{ The single big jump process for in L\'evy Lorentz gas. Panel a,b and c refer to the cases with 0, 1 and 2 reflections respectively. The distance of the scattering points $L$ corresponding to the big jump is in orange, the final position $R$ is in magenta and the total traveled distance in the big jump $v(T-T_w)$ is in blue.}
	\label{LLGjump}
\end{figure}

Let us now introduce our derivation following the reasoning applied for the simple  L\'evy Walk. We assume that the motion of the walker at large distance $R \sim vT$  is determined by a single stochastic event occurring at the crossing time $T_w$. At $T_w$, the walker crosses a scattering point, and this scatterer is followed by a large jump of length $L\sim vT$ where the walker moves ballistically.  Up to time $T_w$, the motion of the walker can be neglected since it is of order $R\sim \ell(T_w)\ll vT$. 
After crossing this long gap, the motion of the walker can be considered deterministic, since
the borders of the gap acts as a perfect reflective walls at least on time scale of order $T$. 
Indeed for a recurrent random walk the probability that the walker is not reflected vanishes at long times. So, the motion, shown in Fig. \ref{LLGjump}, is the following:
up to time $T_w$ the walker remains at the starting point, then it bounces back and forth in the gap of length $L$ for a time $T-T_w$.

First we discuss the property of the crossing time $T_w$.
We call $N_{{\rm eff}}(T_w)$ the number of distinct sites crossed by the walker up to time $T_w$.
$N_{{\rm eff}}(T_w)$ has been studied in \cite{levyrand}, where it is shown that for large enough times $N_{{\rm eff}}(T_w)\sim (T_w/\tau_0)^{1/2}$ if $\alpha>1$ and $N_{{\rm eff}}(T_w) \sim (T_w/\tau_0)^{\alpha/(1+\alpha)}$ if $\alpha<1$ ($\tau_0$ is a suitable time constant). 
 This estimate is obtained taking into account that before entering the large gap the walker has typically moved of $\ell(T)$ and the number of scatterer within a distance $\ell(t)$ is proportional to $\ell(t)$ if $\alpha>1$ and to $\ell(T)^\alpha$ if $\alpha<1$, see \cite{beenakker}.
We define $r_{{\rm eff}}(T_w)$ as the (now time dependent) rate at which the walker crosses scattering sites that have never been reached before, i.e.:
\begin{equation}
r_{{\rm eff}}(T_w)=  \frac{d N_{{\rm eff}}(T_w)}{d T_w} \sim
\begin{dcases}
T_w^{-{1 \over 1+\alpha}} \tau_0^{-{\alpha \over 1+\alpha}} & \mathrm{if}\ 0<\alpha<1 \\
T_w^{-{1 \over 2}} \tau_0^{-{1 \over 2}} & \mathrm{if}\ 1 \leq \alpha  
\label{ptw}
\end{dcases}.
\end{equation}
The value of $\tau_0$ is in general not known since we evaluate $N_{{\rm eff}}$ using a scaling argument which provides only the functional  form of $N_{{\rm eff}}(T_w)$. However, in the final result for $B(R,T)$ $\tau_0$ only determines the value of a global factor which has to be suitably fixed in the comparison with numerical simulations.

We introduce the probability $p_{{\rm tot}}(L,T_w) dT_w dL$ ($L_0/v \ll T_w<T$) at time $T_w$ that the walker enters into a gap of length $L$ never visited before ($L\sim vT$).
Since the distribution of the gap length $\lambda(L)$ 
is time independent we have $p_{{\rm tot}}(L,T_w)=r_{{\rm eff}}(T_w)\cdot \lambda(L)$ where $r_{{\rm eff}}(T_w)dT_w$ is the probability  that at $T_w$ the walker crosses a site never visited before and $\lambda(L)dL$ is the probability  that this site is followed by a gap of length  $L$.  Now we estimate $P(R,T)$ by integrating $ p_{{\rm tot}}(L,T_w) dT_w dL$ over all the paths that at $T$ reach the same distance $R$ and then we change the integration variables from $L$ and $T_w$ to $R$. Once again, it is convenient to study separately the processes performing a different number of reflections
and evaluate the contribution that each process gives to $B(R,T)$. We obtain the scaling form described in Eq. (\ref{single_j2}) with:
	\begin{equation}
	I_{\alpha}(r)= \sum_{n=0}^\infty f_{n,\alpha}(r) 
	\label{gtot}
	\end{equation}
where the functions $f_{n,\alpha}(r)$ describe the processes with $n$ reflections, see Appendix \ref{appendixA2} for details. In particular if the walker does not perform any reflection we have:
\begin{equation}
\begin{array}{cc}
\displaystyle
   f_{0,\alpha}(r) =  \frac{L_0^\alpha}{ v^{1+\alpha} \tau_0^{{\alpha \over 1+\alpha}}}  \frac{1}{(1-r)^{1 \over 1+\alpha} r^\alpha} &  \mathrm{if}\ 0<\alpha<1  \\
   \displaystyle
   f_{0,\alpha}(r) = \frac{L_0^\alpha}{ v^{1+\alpha} \tau_0^{{1 \over 2}}} \frac{1}{(1-r)^{1 \over 2} r^\alpha} 
   & \mathrm{if}\ 1 \leq \alpha  
\end{array} 
\label{f0}
\end{equation}
while for an odd number and an even number $n>0$ of reflections we get respectively:
\begin{widetext}
\begin{equation}
\begin{array}{cc}
\displaystyle
f_{n,\alpha}(r) = \frac{L_0^\alpha (n+1)^\alpha \alpha}{v^{1+\alpha} \tau_0^{{\alpha \over 1+\alpha}}}  \theta(1-nr) \int_{0}^{1-nr} \frac{dt_w}{t_w^{{1 \over 1+\alpha}} (1+r-t_w)^{1+\alpha}}  & \mathrm{if}\ 0<\alpha<1                 \\
\displaystyle
f_{n,\alpha}(r) =  \frac{L_0^\alpha (n+1)^\alpha \alpha}{v^{1+\alpha} \tau_0^{{1 \over 2}}}  \theta(1- nr) \int_{0}^{1-nr} \frac{dt_w}{t_w^{{1 \over 2}} (1+r-t_w)^{1+\alpha}}
        & \mathrm{if}\ 1 \leq \alpha  
\label{gnpr}
\end{array}
\mathrm{when\ } n\ \mathrm{is\ odd}
\end{equation}
\begin{equation}
\begin{array}{cc}
\displaystyle
f_{n,\alpha}(r) =  \frac{L_0^\alpha  n^\alpha \alpha}{v^{1+\alpha} \tau_0^{{\alpha \over 1+\alpha}}}  \theta(1-(n+1)r) \int_{0}^{1-(n+1)r} \frac{dt_w}{t_w^{{1 \over 1+\alpha}} (1-r-t_w)^{1+\alpha} } & \mathrm{if}\ 0<\alpha<1                 \\
\displaystyle
f_{n,\alpha}(r) =  \frac{L_0^\alpha  n^\alpha \alpha}{v^{1+\alpha} \tau_0^{{1 \over 2}}} \theta(1-(n+1)r) \int_{0}^{1-(n+1)r} \frac{dt_w}{t_w^{{1 \over 2}} (1-r-t_w)^{1+\alpha}} { }       & \mathrm{if}\ 1 \leq \alpha
\label{gndr}
\end{array}
\mathrm{when\ } n\ \mathrm{is\ even}.
\end{equation}
\end{widetext}
The integral defining $f_{n,\alpha}(r)$ can be evaluated numerically.
Due to the $\theta$-functions in $f_{n,\alpha}(r)$, for any value of $r$ only a finite number of terms gives a non zero contribution to the sum in (\ref{gtot}). In particular, as $r \to 0$ the number of terms must be increased, while for $r>1/3$ only two terms are needed. This means that $I_{\alpha}(r)$ is non analytic (the derivatives are discontinuous) when 
\begin{equation}
r=\frac{1}{2m+1} \ \ {\rm with}\ \  m=1,2,\dots
\end{equation}
This is a consequence of the non analytic dynamics at the reflections in $0$ and $L$.

In Appendix \ref{appendixA2} we also analyze the behavior of $I_{\alpha}(r)$ at small $r$ showing that  $I_{\alpha}(r) \sim r^{-(1+\alpha)}$ for $r\to 0$. This means that $I_{\alpha}(r)$ is not an integrable function
in $r=0$ and therefore  it is an infinite density as in previous cases.

\subsection{Numerical results}

Let us now compare the analytical results with numerical simulations. In Fig. \ref{scalp} the PDF is rescaled according to Eq. (\ref{gtot}) and the theoretical scaling function $I_{\alpha}(r)$ is plotted with a thick magenta line.
Here we used the same simulation data of Fig. \ref{scalSH}, introducing only a different scaling procedure.  $I_{\alpha}(r)$ has been evaluated summing up to $100$ terms in Eq. (\ref{gtot}) so that the numerical error on the analytic result is negligible at least for $r>.01$. The curves scale quite well and they collapse on the predicted function. Clearly numerical results are closer to the analytical prediction for large times and large $r$. Indeed our result is exact in the limit $R\gg\ell(T)$ and $T\gg\ell(T)$. The figure shows the non analytic behavior of $I_{\alpha}(r)$ with a discontinuity in its derivative for $r=\frac{1}{2m+1}$ with $m=1,2,\dots$. Also these non-analyticities in simulations are observed only in the long asymptotic regime when the reflection time is negligible with respect to the evolution time, giving rise to instantaneous non-analytic reflections. 

We also remark that at small $\alpha$ (i.e. $\alpha<1$) the numerical simulations converge to the scaling function at very long times, indeed in this case $\ell(T)$ grows faster and the condition $\ell(T)\ll vT$ is realized at larger $T$. However for small $\alpha$ large time simulations can be numerically afforded quite easily, as the computational times grows with the number of scattering events and not with $T$. On the other hand, for large $\alpha$ (i.e. $\alpha>2$) the curves converge at small times but simulations are very demanding: they require an average over a huge number of disorder realizations, since in this case the ballistic stretch with $L\sim vT$ are extremely rare events (see the number of realizations for data at $\alpha=2.2$ which are still very noisy.).

\begin{figure*}
	\includegraphics[width=0.45\textwidth]{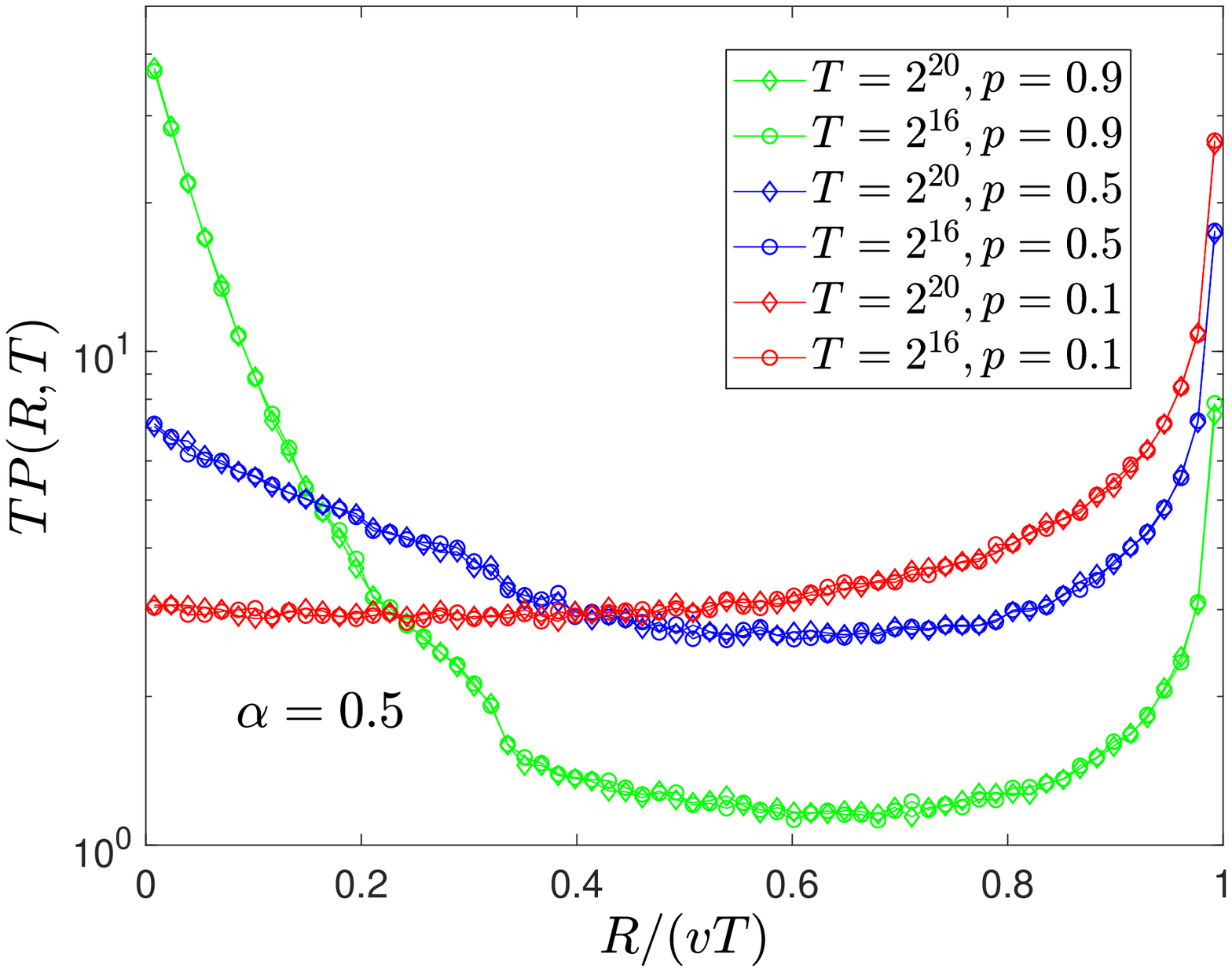}
	\includegraphics[width=0.45\textwidth]{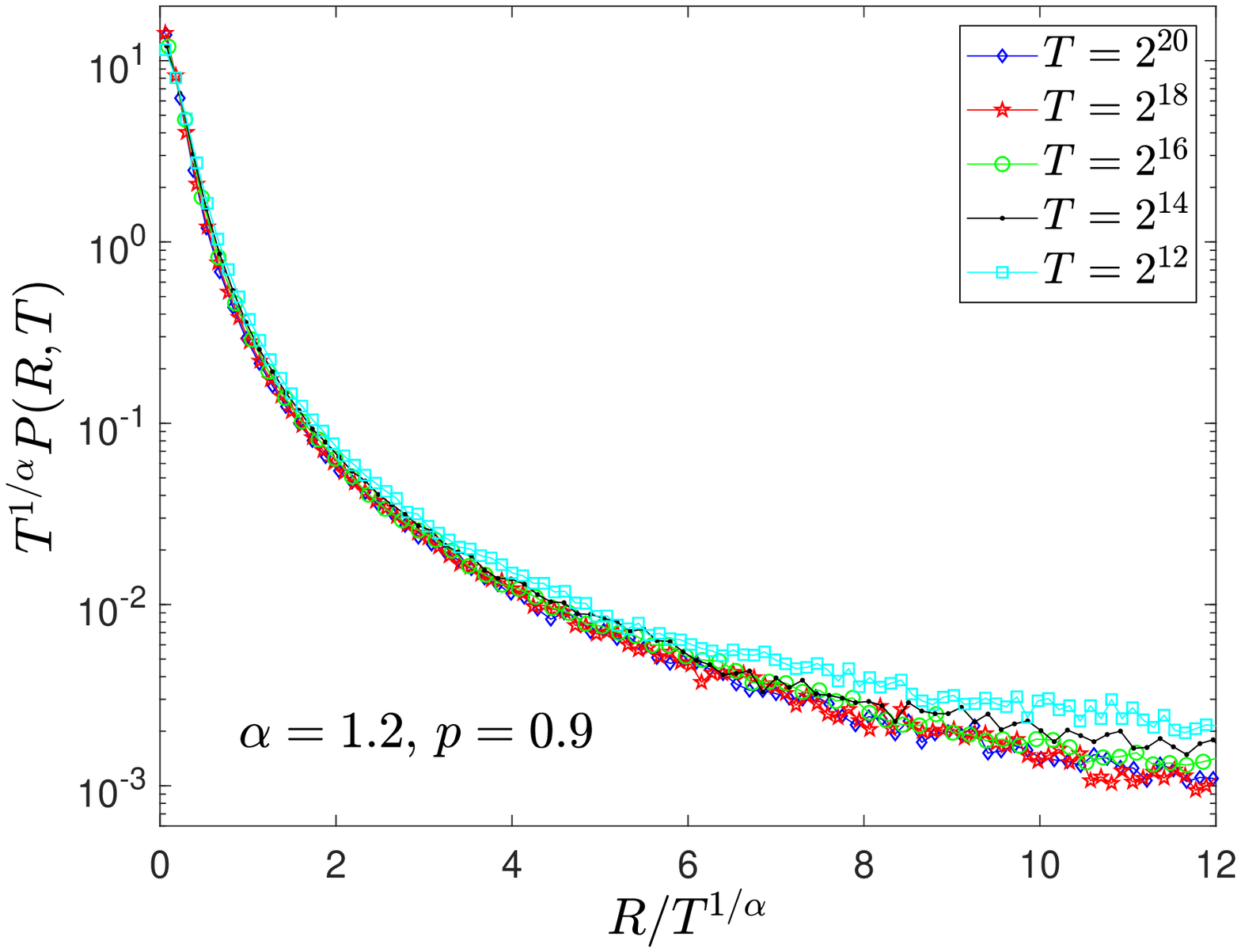}
	\includegraphics[width=0.45\textwidth]{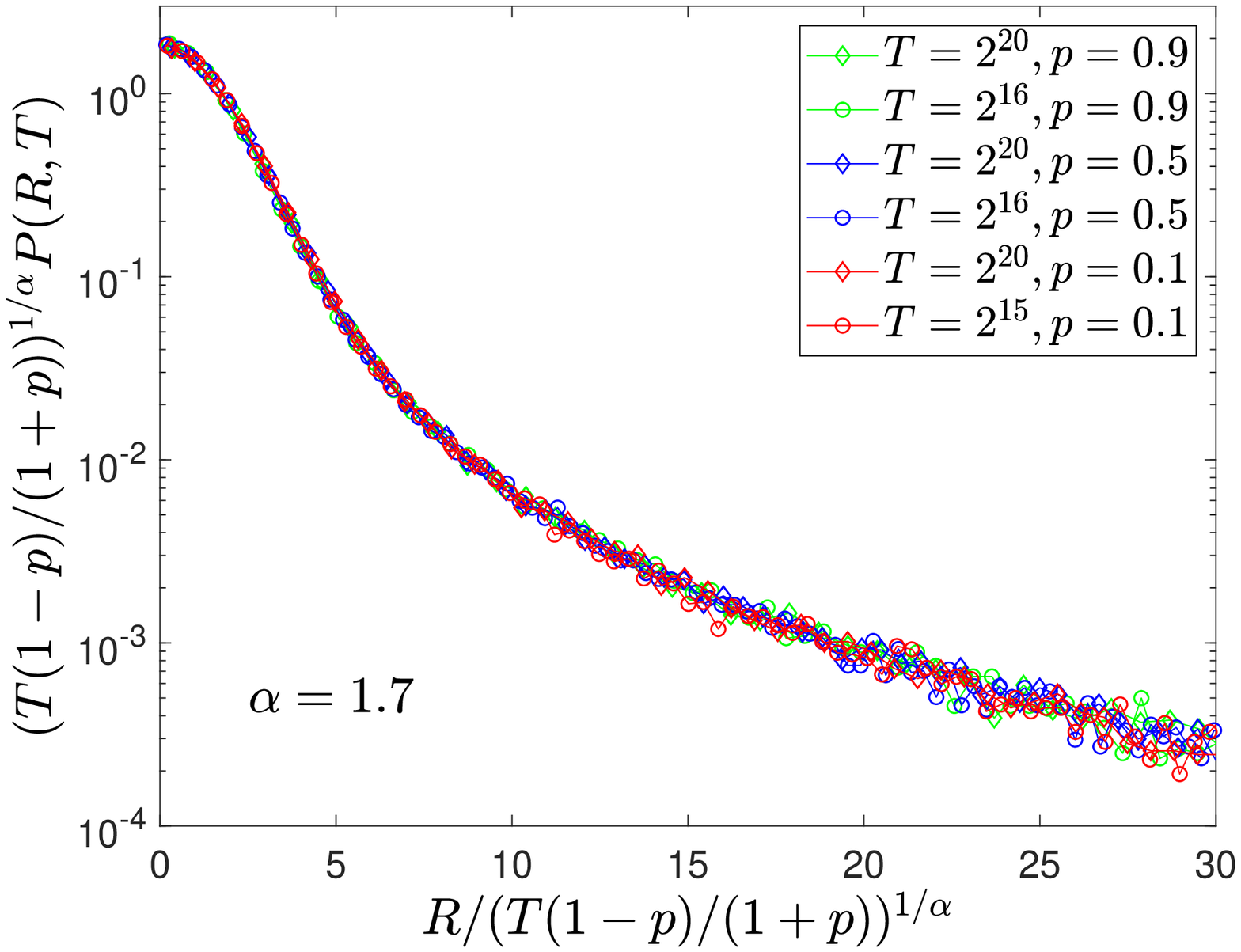}
	\includegraphics[width=0.45\textwidth]{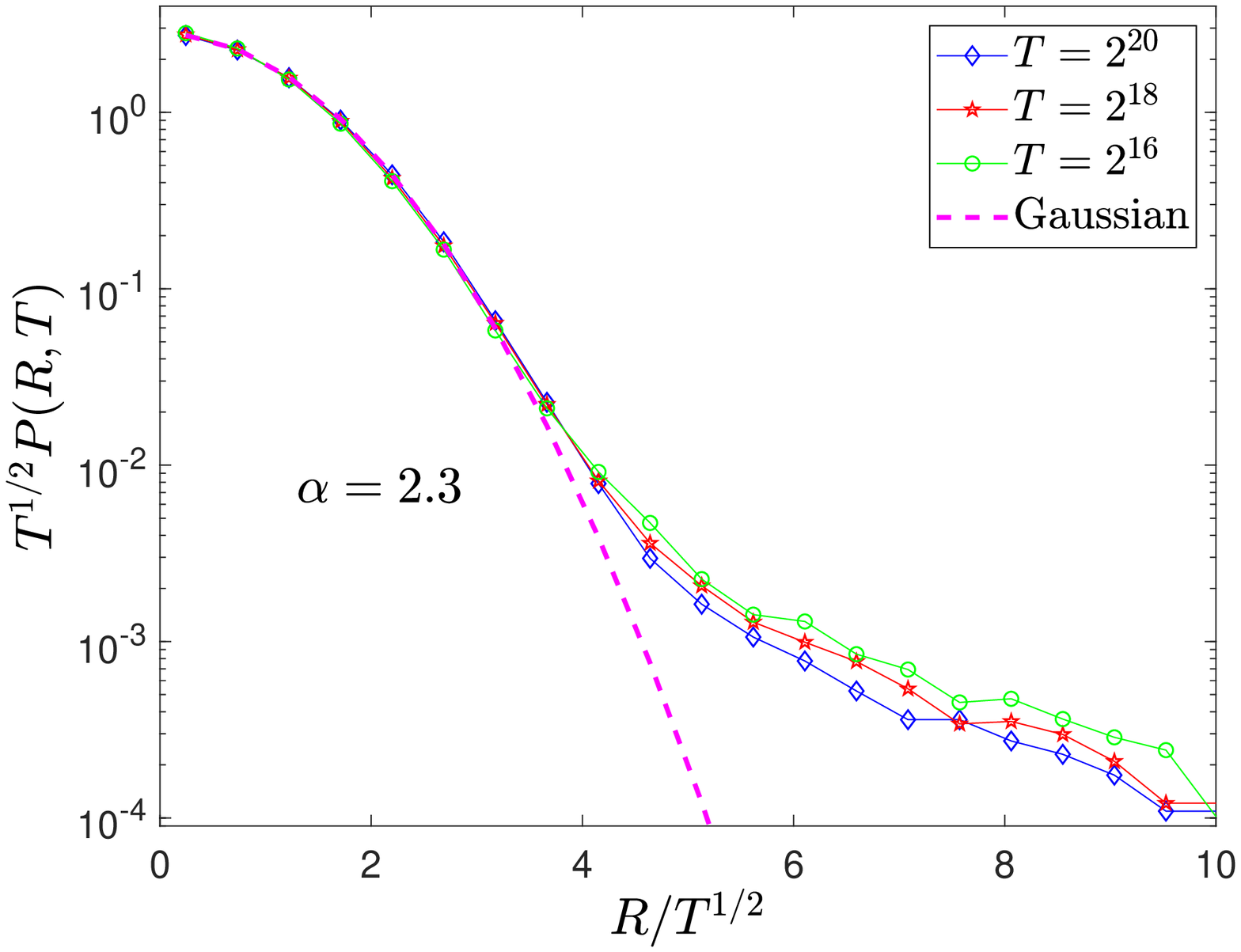}
	\caption{(Color on-line) Scaling at short distances for the PDF in the correlated L\'evy walk model. The case $\alpha=1.7$ shows that for $\alpha>1$ a suitable rescaling of space and time allows to rescale data relevant to different values of $p$, while the case $\alpha=0.5$ clarifies that for $\alpha<1$ the scaling function is not universal depending both on $\alpha$ and $p$. Finally, for $\alpha>2$ the standard Gaussian behavior is recovered due to central limit theorem.}
	\label{scalSHc}
\end{figure*}

In Fig. \ref{scalpLog} using a logarithmic scale we show that $I_{\alpha}(r) \sim r^{-(1+\alpha)}$ at small $R/(vT)$, i.e. one has to consider to be valid  both the regimes $R\gg \ell(T)$ and $ R \ll vT$. We notice that in simulations the time scales are too small to sample this power law regime.
 
\section{Correlated L\'evy walks}
\label{correl}

The L\'evy-Lorentz gas can be considered a peculiar example of a correlated L\'evy walk. Indeed, jumps in the spatial region which has not been yet reached by the walker are renewals of the motion, i.e. one can say that step lengths are randomly extracted from $\lambda(L)$. On the contrary, in regions which have been already visited by the walker, the motion is strongly correlated. Indeed, in the same framework, the length of the jumps is fixed by the previous evolution of the walker which determines the position of the scattering points. Therefore, one can argue that the big jump argument can be applied in a wide range of correlated random walks with memory characterized by sub-exponential big jumps. Let us introduce an example clarifying this possibility. 

We consider a random walk that at each step covers with probability $(1-p)/2$  at velocity $v$ a distance $L$ extracted from $\lambda(L)$ defined in Eq. (\ref{lambdaL}); 
with the same probability it covers the same distance $L$ but at velocity $-v$; finally, with probability $p$
the walker makes a jump of the same length of the jump in the previous step, but it moves with opposite velocity, i.e. its reflected to the starting point of the previous step. This dynamical rule gives rise to a correlation in the motion of the walkers and an analytic study of the PDF $P(R,T)$ is non trivial due to memory effects. However,  correlations decays exponentially with the number of steps as $p^n$, therefore the universal behavior is the same of the standard L\'evy walks. In particular, as we show in Appendix \ref{appendixB} at short distances for $\alpha>1$ one recovers the same behavior of standard L\'evy walk provided that time is rescaled by a factor $(1-p)$ and space is rescaled  
by $(1+p)^{1/2}$ for $\alpha>2$ and by $(1+p)^{1/\alpha}$ for $1<\alpha<2$. Numerical simulations in Fig. \ref{scalSHc} 
show indeed that for $\alpha>2$ the PDF is Gaussian and it has a diffusive scaling. For $1<\alpha<2$ we show that after rescaling space and time the scaling length becomes $\ell(T)\sim ((1-p)/(1+p)T)^{1/\alpha}$ so that the PDF scales according to a typical L\'evy behavior $P(R,T)=\ell(T)^{-1}L_\alpha(R/\ell(T))$ where $L_\alpha(\cdot)$ is this is the symmetric stable L\'evy density independently of $p$. Finally for $\alpha<1$ the ballistic motion dominates and $P(R,T)=T^{-1}f_{\alpha,p}(R/T)$ where $f_{\alpha,p}(T)$ is a non universal scaling function depending both on $\alpha$ and $p$.

\begin{figure}
	\includegraphics[width=0.45\textwidth]{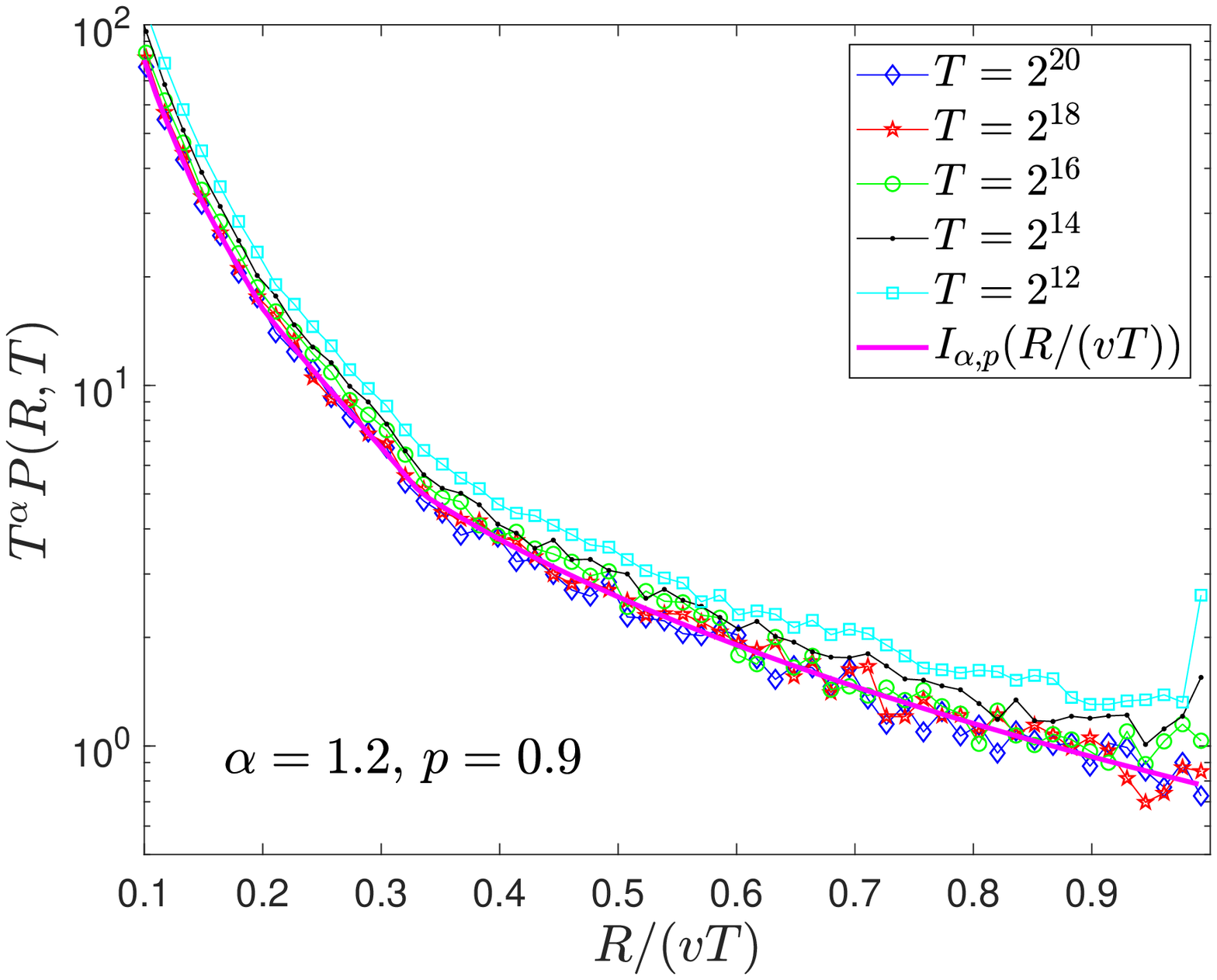}
	\includegraphics[width=0.45\textwidth]{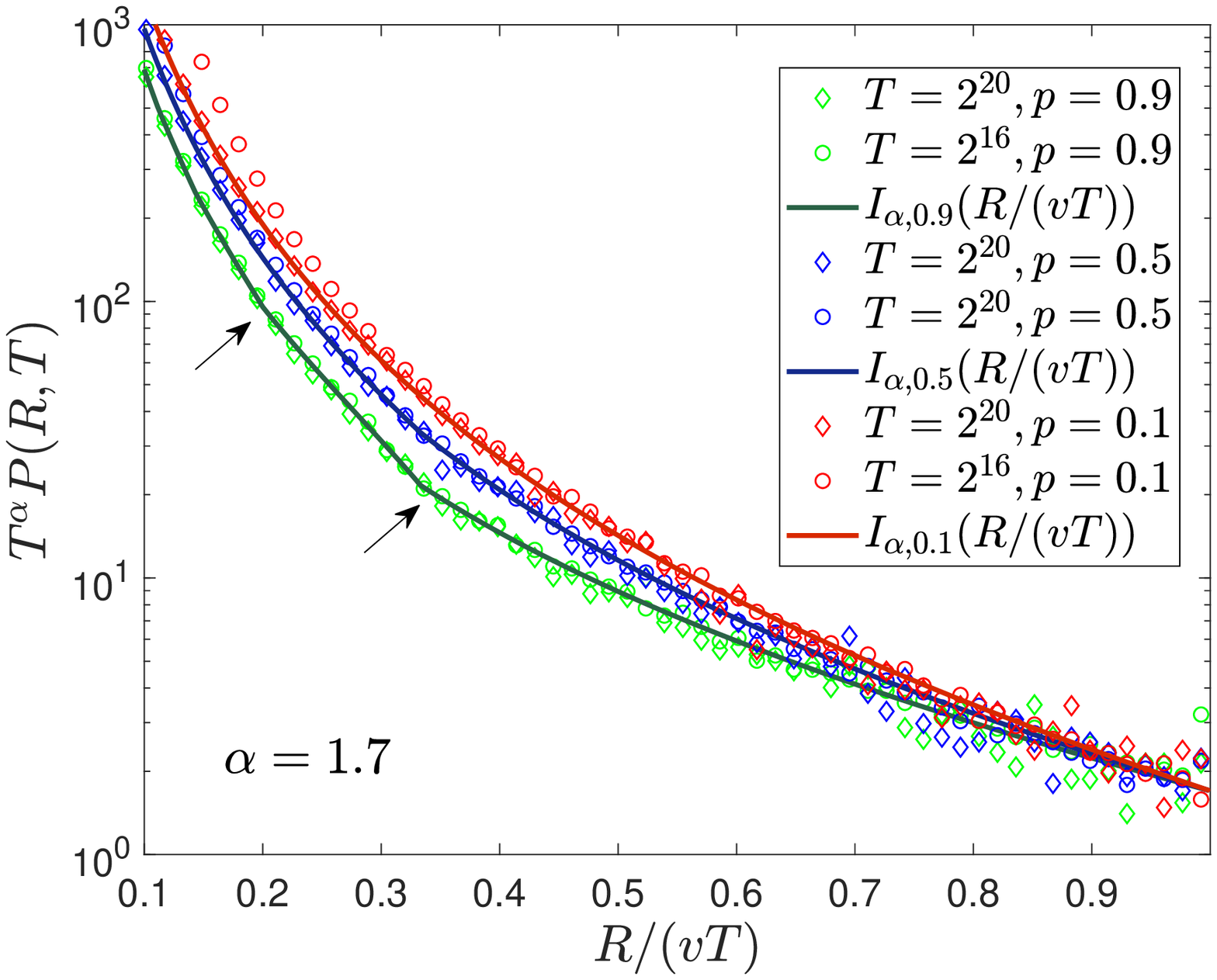}
	\includegraphics[width=0.45\textwidth]{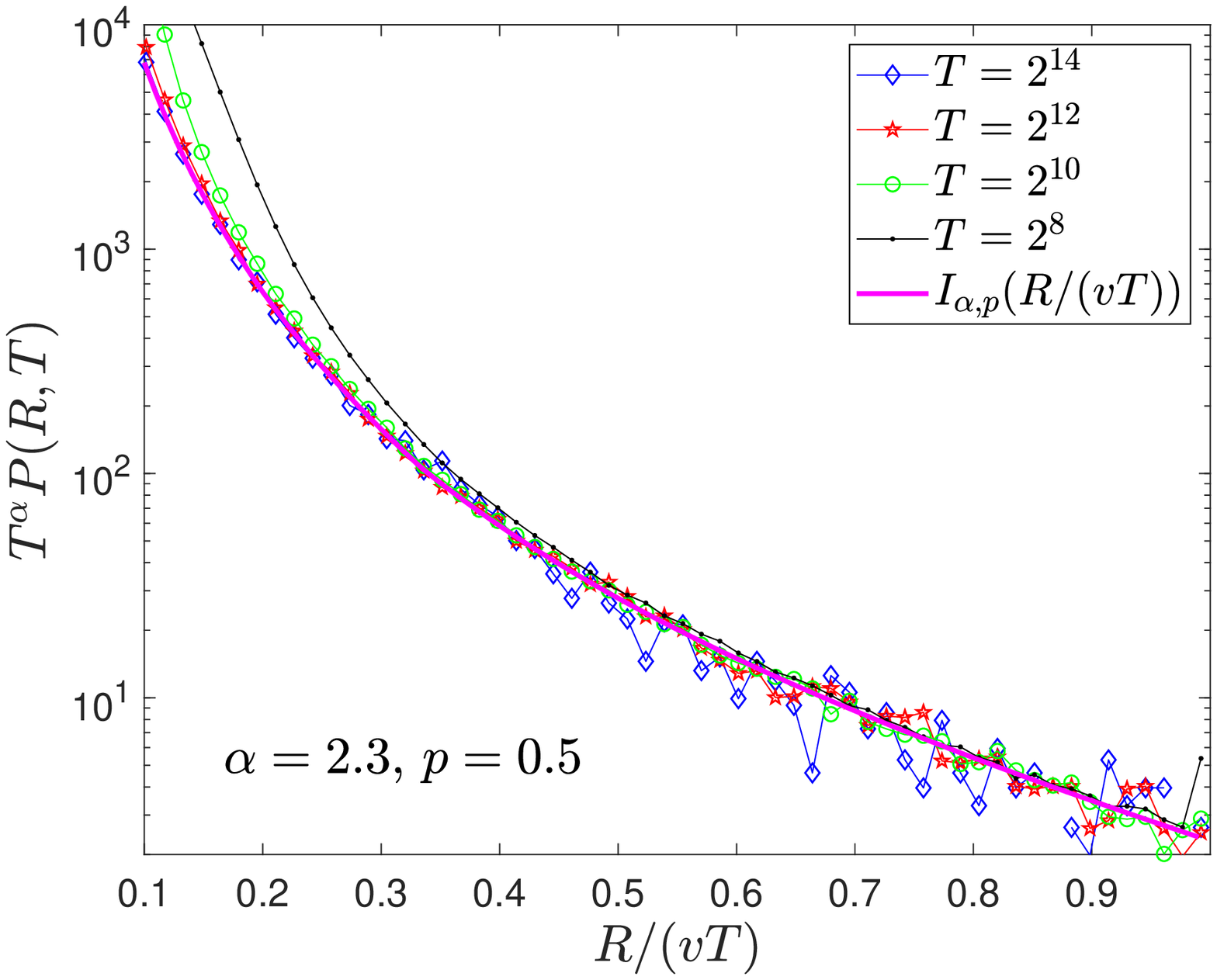}
	\caption{(Color on-line) Scaling at large distances for the PDF in the correlated L\'evy walk model. Data are obtained considering 
		$5 \cdot 10^6$ realization of the process for $\alpha=1.2$ and $\alpha=1.7$; while $2\cdot 10^8$ realizations have been considered for $\alpha=2.3$.}
	\label{scalc}
\end{figure} 
 
Let us show that, for $\alpha>1$, one can use the single jump approach to calculate the  behavior of the PDF at large $R$ (i.e. $R\sim T \gg\ell(t)$). We introduce 
the probability at time $T_w$  $p_{{\rm tot}}(L,T_w) dL dT_w$ of making a jump of length $L$, with $L_0/v \ll T_w<T$. For $\alpha>1$ the average duration of a step $\langle \tau \rangle$ is finite, so that the jump rate at $T_w$ is $1/ \langle \tau \rangle$ independently of $T_w$ and we have: $p_{{\rm tot}}(L,T_w)= (1-p) \lambda(L)/\langle \tau \rangle$; where the factor $(1-p)$ takes into account the reflection probability.
Also in this case we have to consider all the path driving the walker in $R$ at time $T$.  In this framework, apart the big jump, all other steps can be neglected. Therefore, the effective motion can be described as follows: the walker at $T_w$ performs a jump of length $L$ with probability $p_{{\rm tot}}(L,T_w)dL dT_w$.  At the end of this jump the walker is reflected with probability $p$ or it remains stuck in $L$ with probability $1-p$; after the reflection the walker returns to the origin where again it can be reflected with probability $p$ or it halts with probability $1-p$, and so on. I.e. the origin of the walker and the point at distance $L$ act as scattering points
where the walker is reflected or absorbed with probability $p$ or $1-p$ respectively.  In analogy with L\'evy walks and L\'evy Lorentz gas one can evaluate  the contribution of the different processes to the PDF $P(R,T)$. In this case, once $T_w$ and $L$ are fixed, the final position at $T$ is not fully determined and one have to consider the probabilities of different paths depending on $p$. In particular, if $L>T-T_w$ the walker is still covering the big jump at $T$ and the contribution of this process is:
\begin{equation}
{\tilde B}_0(r,T)= 
\frac{1}{T^{\alpha}} \frac{(1-p)L_0^\alpha}{ v^{1+\alpha} \langle \tau \rangle} \frac{1}{r^\alpha} 
\label{c0r}
\end{equation}
where $r=R/vT$. The probability to be in motion at the position $R$ after $n$ reflections is
\begin{eqnarray}
{\tilde B}^r_n(r,T) & = &
p^n \frac{1}{T^{\alpha}} \frac{(1-p)L_0^\alpha (n+1)^\alpha}{ v^{1+\alpha} \langle \tau \rangle}
\cdot \nonumber\\ 
&\ & \theta(1-nr)\cdot \left( \frac{1}{((n+1)r)^\alpha}-\frac{1}{1+r}\right) 
\label{c1r}
\end{eqnarray}
if $n$ is odd and 
\begin{eqnarray}
{\tilde B}^r_n(r,T)& = & 
p^n \frac{1}{T^{\alpha}} \frac{(1-p)L_0^\alpha n^\alpha}{ v^{1+\alpha} \langle \tau \rangle}
\cdot \nonumber \\
&\ & \theta(1-(n+1)r)\cdot \left( \frac{1}{(nr)^\alpha}-\frac{1}{1-r}\right) 
\label{c2r}
\end{eqnarray}
if $n$ is even. Eqs. (\ref{c1r},\ref{c2r}) are analogous to Eqs. (\ref{gnpr},\ref{gndr}) since both describes reflection of the walker in a gap of length $L$ integrated over all possible lengths. In this second case, integrals can be evaluated exactly since the jump rate is independent of $T_w$ providing a more simple integrand function.
Moreover, here we have the factor $p^n$ representing the probability that $n$ reflections occur during the evolution and the walker does not halt before time $T$.
Finally one has to consider the contributions when the walker remain stuck in $L$ or in $0$ at the $n$-th scattering event. For $n$ odd (i.e. the scattering occurs in $L$) we have:
\begin{equation}
{\tilde B}^s_n(r,T)= 
(1-p) p^{n-1} \frac{1}{T^{\alpha}} \frac{(1-p)L_0^\alpha \alpha}{ v^{1+\alpha} \langle \tau \rangle}
\cdot \theta(1-nr) \frac{1-nr}{r^{\alpha+1}} 
\label{c3r}
\end{equation}
Eq. (\ref{c3r}) is analogous to Eq. (\ref{g1w}) for L\'evy walks since in both cases they describes a walker that halts in $L=R$. Clearly in L\'evy walks reflections are not possible so $p=0$ and $n=1$. The factor $(1-p) p^{n-1}$ is the probability that the walker is reflected for the first $n-1$ scattering events and absorbed at the $n$-th event. The $\theta$-function takes into account that if there are $n$ scattering the distance cannot be larger of $vT/n$. Finally if the walker halts at the $n$-th scattering event with $n$ even, i.e. it halts in the origin, the process does not give any contribution to $P(R,T)$ since it is not a big jump. This means ${\tilde B}^s_n(r,T)=0$ for $n$ even.
The rescaled PDF $P(R,T)$ can then be evaluated by summing Eqs. (\ref{c0r}-\ref{c3r}) i.e.:
\begin{eqnarray}
& & P(R,T) = \frac{1}{T^{\alpha}} I_{\alpha,p}\left(\frac{R}{vT},T\right) =\\
& & ={\tilde B}_0\left(\frac{R}{vT},T\right)+\sum_{n=1}^{\infty}{\tilde B}^r_n\left(\frac{R}{vT},T\right) +{\tilde B}^s_n\left(\frac{R}{vT},T\right)\nonumber
\label{c4r}
\end{eqnarray}
$I_{\alpha,p}(r)$ is an infinite density since it diverges for $r\to 0$ as $r^{-1-\alpha}$.
We notice that $I_{\alpha,p}(r)$ is non-analytic for $r=\frac{1}{2m+1}$ with $m=1,2,\dots$ i.e. the same values of the L\'evy Lorentz gas.
In Fig. \ref{scalc} we compare the analytical results with numerical simulations finding a good agreement. In the case $\alpha=1.7$ we show that  $I_{\alpha,p}(r)$ depends explicitly on the parameter $p$; therefore it is not an universal scaling function.
For $\alpha=1.7$ and $p=0.9$ we also show the non-analytic points of 
$I_{\alpha,p}(r)$ which are in general less pronounced than in L\'evy Lorentz gas. Since this points are originated, also in this dynamics, by the reflections they become more visible when $p$ is closed to $1$ i.e. when these reflections are more probable.
As in the previous cases, at large $\alpha$, since rare events are always less probable, simulations become computationally very demanding (see the huge number of realizations) .

\section{Discussion and Open Problems}

The single big jump principle is a statement about the origin of rare events in fat tailed processes: it allows to identify the 
mechanisms that lead to the rare fluctuations. Then, we can use different techniques to calculate the contribution of the big jump 
and we showed that several approaches can be applied. 
For the L\'evy walks, we used extensions of the Frechet Law, Eq.(15),  and then a heuristic argument, 
that matches the results for the calculation of rare events obtained from the moment generating function and infinite densities approach. 
For more complex cases, we applied a rate approach, that consists in splitting the problem in short time and long time dynamics.
We first calculated the PDF of performing a jump of length much larger than the characteristic length of the process, by determining the effective rate $r_{{\rm eff}}$ at which jumps are made. Then we summed over all paths that reached that distance.  In this way, the short distance dynamics is condensed in the rate, while the role of correlations is resumed in the sum over the different paths.  Within this scheme, the big jump principle allows for a direct physical interpretation of the processes reaching large distances, and at the same time it provides an effective tool for calculations. Interestingly, for the L\'evy Lorentz gas the short time dynamics is unknown, however an estimate of the rate  $r_{{\rm eff}}$ is enough to apply the principle and calculate the non trivial shape  of the PDF at large $L$.  

For the cases under study in this work, we observe a non-uniform
description of the PDF: typical and rare events do not scale with time in the same manner. More precisely, from central limit theorem the bulk of the PDF of random walks is described in term of a single characteristic scaling length which grows with time $T$ as $\ell(T)\sim T^\gamma $, where $\gamma$ smaller, equal or larger than $1/2$ stands for sub, normal and super diffusion respectively.
The study of the far tail requires a different picture since the spatio-temporal correlations in the single step give rise to a new scaling length, which determines the behavior of the single big jump regime. We have shown that the competition between these two scaling lengths produces a dual scaling of the moments, which has been termed by Vulpiani and co-workers as strong anomalous diffusion \cite{castiglione}. This means that for $q>q_c$ we have $\langle |x(T)|^q\rangle \not \sim \ell(T)^q$: large moments are not determined by the bulk distribution  but by its far tail. 
This in turn implies that the single jump is needed, at least in some cases, for the investigation of the mean square displacement of the spreading process,  which is easily considered the main quantifier of diffusion.  Since strong anomalous diffusion is observed in a wide range of systems, we believe the principle of large jump has wide spread applications \cite{Xiong,Holz,Krapf,Cagnetta,deAnna}.

Further, for the simpler models under investigation, i.e L\'evy  walks and cold atom system, the single big jump allows a new insight on non-normalised covariant densities \cite{Eli1,Eli2}. As mentioned the far tails  of the density, obtained from the big jump principle, controls the behavior of moments of interest. These in turn  can be calculated from certain non-normalised densities, so the principle used here explains precisely the physical origin of these mathematical tools: they stem from one big jump. 

Interestingly, in the physical literature the single big jump principle has also been related to condensation in probability space: the probability of 
the sum $x=\sum_{i=1} ^N x_i$ {\em condenses} to the probability of a single variable \cite{Maj1b,Maj2,filias,corberi}, that is the maximum value of $x_i$.  
In condensation problems, the phenomenon where a large macroscopic portion of particles occupies a single state is well known. For example distribution
of masses occupying lattice sites  in a system can contain, in certain conditions, one region where a vast majority of mass is located, while other regions
are sparsely populated. In practice, the summands we consider do not have to be displacements, instead they can be  masses condensing on a lattice, so that $N$ would be the system size,
or it could be  energy etc. Therefore, the approach could be extended to
systems interacting  with a reservoir of particles/energy, so that the total mass/energy/etc can fluctuate, providing therefore a general background for large fluctuations estimates in different frameworks.

There are still many open problems, and we briefly discuss some of them.

\begin{itemize}

\item[1.]
For the case of summation  of IID random variables
 the principle of big jump  works for any $N$.
It is therefore natural to ask, for the more general case,
does the principle hold for all times? 
For intermediate or short  times the analytical predictions presented in this manuscript 
 are  not valid, since we have used the long time limit. 
Clearly this does  not mean that the principle is not valid at all times,
but rather that the analytical formulas are not elegant or attainable
at short times. 
For short
times the initial conditions play a special role, however this holds both for
the maximal displacement and the total displacement. We leave it for
future work to check the principle, and this could be done numerically,
for example for the L\'evy walk, by a calculation of the far tail of the total
displacement and comparing with  the statistics of the biggest jump. 
Our first  simulations presented  in Fig. 3 show that in the tail the two distributions
agree for all times we have considered.

\item[2.] For the L\'evy walk, we focused our attention on the case when $\alpha>1$. Hence
the mean time between collision events or zero crossings 
 for all the  Models was finite. 
It is expected that the big jump principle holds also for
$0<\alpha<1$, however the explicit formulas for the far tails of the density
need to be analyzed with different tools than those presented here.
Work in this direction is required to further establish the generality 
of the principle.

\item[3.] Bi-scaling of moments was observed for tracer particles diffusing in the cell \cite{Gal}. Can we use the principle in the context of diffusion of particles in that case? 
Active transport mediated by ATP, i.e. the pumping of energy
to the system is responsible for this behavior. From data analysis one can see many small displacement of the tracer particle,
and a few large jump events. Importantly when removing the large jumps from the data set, one can
see mono-scaling, thus the observed strong anomalous diffusion is clearly related to large jumps (we cannot say if a few
or a single jump). We believe that rare events in this system will be described by the big jump principle.
In principle this is easy to check, considering the distribution of the sum of displacement  of the particle and comparing
it to the distribution of the maximum of the displacements. 
In turn the characterization of the tails of density of particles diffusing in the cell,
is clearly important, since this helps in the understanding of active transport, and also since these
rare events are important for the exploration of the cell environment. Imagine a particle 
diffusing some time in the cell, looking for a target (a reaction center): if the target is not found with in some time interval
it might be beneficial for the particle
to relocate and start its search yet again. However, the efficiency of such search is controlled by the large jumps, and
hence quantification and verification of the big jump principle, in the context of diffusion of molecules
in the cell, might turn out to be important.

\item[4.] 
The principle of biggest jump is related to the calculation of distribution of forces in long ranged interacting systems, governed by Coloumb or gravitational fields, like plasma and astrophysics \cite{plas1,plas2}. 
For example the distribution of the force acting  on a unit mass (or charge) embedded in a sea of masses  (or charges).  Considering  the former case, with the masses uniformly distributed in space, 
the force acting on
a single element is a sum many forces, and for long range gravitational or Coulomb forces
this leads to the well known Holtsmark distribution for the forces \cite{Holtsmark,Ruffo,Chandrasekar,Chavanis,Chavanis2}. 
This problem in different variants appears in many systems \cite{Kuvsh}. One can argue that the influence
of the nearest neighbor is most important, namely instead of summing over all forces,
we need to consider only the nearest neighbour, and this is certainly in spirit of the big
jump principle.  When the masses (charges) are uniformly distributed, the problem is related
to summation of IID random variables, and indeed L\'evy central limit theorem is known to
describe the statistics of this problem. It would be interesting to check the fluctuations
of the forces, in the limit of large forces, based on the biggest jump/force for cases
where the masses are arranged more realistically in space. 
On an operational level, the big jump principle suggests that, for the sake of the large
fluctuations of the forces, we need only partial information on the system, namely 
the random distance to nearest neighbour.

\item[5.] Of course an interesting topic is the estimation of the big jump statistics from data, for example when we follow a trajectory of a single molecule in the cell,  or when we sample 
the trajectory  with a given rate, for finite time, and the number of trajectories might not be very large.  Another important step is how to extract the big jump from time series of events, 
and this could be done by  analyzing a correlation plot, analogous to the one presented in Fig. 4 of Section II.B  for the L\'evy walk. These sampling effects should  be further investigated.

\end{itemize}

\section{Perspectives}

The Big Jump principle applied to physical modelling is an extremely powerful tool that can be used to estimate
probabilities of rare events in a wide range of interesting problems, in the presence of fat tailed distributions.
The principle was  reformulated, extended and tested  far beyond the case of a sum of IID random variables.  
It was extended to correlated processes, continuous processes, systems with quenched disorder and processes with a finite upper 
speed, all of which lead to important modifications of the principle of big jump in its original  form. 
Simply said, the mentioned effects make the sub-exponential tail far from trivial, while the case
of IID random variables leads to a simple power law tail, which is not applicable as we have shown. 

Given the fact that extreme events are model specific, we find it very encouraging that we can at all formulate a general principle to describe their behaviour. Thus while the
shape of the packet density in its far tails varies from one model to the other, all of them are described by the statistics of the biggest jump.
At the same time, this is a warning sign to any one dealing with predictions of rare events. If one events is controlling the statistics of extremes,
we might understand better the inherent difficulties in predictions, but at the same time understand how to quantify these extremes better. For example, consider
the accumulated rain fall in say one month in some region. The accumulated rain fall is important for example if we plan a reservoir which holds the water, or
if the total rain fall per fixed time (say one month) is of critical importance. There are strong experimental evidences that the statistics of rain falls could be
 a test bed for our theory \cite{rain}.
First, with the principle at hand we can use records, or models to see if the
principle works. For example comparing the total rain fall within a month to the maximum of rainfall per day. 
Then we may determine if a systems behaviour is close or not to the principle of big jump.  At least in principle,  policy takers could reach educated decisions, as the answer
to the question:  do we get prepared to one big event (one day of massive water fall) or do we prepare
for  many accumulated  events, could be tackled with wisdom. 

While this demands further work, our theory is already shedding light on important physical
processes, beyond the IID case  \cite{Lucilla1}. We have recently  worked on models of active particles propagation  and contamination spreading in the field of Hydrology. 
Here deep traps in the spirit of the trap model and continuous time random walks are extensively used.
In this case all the spatial jumps are small, and narrowly distributed, so there is no spatial big jump. However, one can apply our principle to the concept of time jumps, namely look for the longest stalling time in  these processes. This, as we will show in a later publication, gives insight on the mentioned processes, shedding new light on the far tails of the spreading phenomenon. Thus while we dealt with models where the particles are always in motion, and never trapped, we can extend our work to
model trapping events, where the motion  is typically considered slower than normal.  These in turn are widely applicable, in a vast number of systems, hence we know that the principle of big jump can be a turning point in the analysis of rare events in many systems.

\section{Acknowledgement}

This research was partially funded by the Israel Science Foundation (EB) and by the CSEIA Foundation (RB). EB thanks David Kessler, and Erez Aghion, for discussions.



\appendix

\section{Single big jump in the L\'evy Lorentz gas}
\label{appendixA2}

Let us call $B_n(R,T)$ the contribution to $B(R,T)$, which is obtained
integrating $p_{{\rm tot}}(L,T_w)$ over all the processes that in a time $T$ arrives in $R$ after $n$ reflections .
If $L>v(T-T_w)$ no reflection occurs and $R=v(T-T_w)$, i.e. $L>R$, $T_w=T-R/v$ and $dT_w=dR/v$. Clearly all the jumps of length $L>R$ contribute to the process ending in $R$, so $B_0(R,T)$ is:
\begin{eqnarray}
B_0(R,T) dR & =& r_{{\rm eff}}(T-R/v) dR/v \int_{R}^{\infty}\lambda(L) dL \\
&=&
\begin{dcases}
\frac{dR \left(\frac{L_0}{R}\right)^\alpha}{v \tau_0^{{\alpha \over 1+\alpha}} (T-R/v)^{{1 \over 1+\alpha}}}  & \mathrm{if}\ 0<\alpha<1 \\
\frac{dR \left(\frac{L_0}{R}\right)^\alpha}{ v \tau_0^{{1 \over 2}} (T-R/v)^{{1 \over 2}}}   & \mathrm{if}\ 1 \leq \alpha  
\label{g0}
\end{dcases}.\nonumber
\end{eqnarray}
Introducing the rescaled adimensional variables $r=R/(vT)$ ($0<r<1$) and the rescaled function $\tilde B_0$, we get:
\begin{equation}
{\tilde B}_0(r,T)= 
\begin{dcases}
\frac{1}{T^{\frac{1+\alpha+\alpha^2}{1+\alpha}}} f_{0,\alpha}(r) & \mathrm{if}\ 0<\alpha<1 \\
\frac{1}{T^{\frac{1}{2}+\alpha}} f_{0,\alpha}(r) & \mathrm{if}\ 1 \leq \alpha  
\label{g0r}
\end{dcases}
\end{equation}
where $f_{0,\alpha}(r)$ is given by Eq. (\ref{f0}).

If $L<v(T-T_w)$, the walker is reflected in $L$ 
then it moves in the opposite direction and if $v(T-T_w)<2L$ the second reflection in $R=0$ does not occur before $T$.
Let us call $D$ the distance covered by the walker after the reflection in $L$. We have $L+D=v(T-T_w)$ and $R=L-D$, so we get $2L=R+v(T-T_w)$. Imposing $v(T-T_w)/2<L<v(T-T_w)$ we get 
$v(T-T_w)/2<R+(T-T_w)/2<v(T-T_w)$. The first inequality is trivially satisfied while the second gives the condition $T_w<T-R/v$.
To get the probability of reaching $R$, we can integrate over the processes that for different $T_w$ arrive at the same position: 
\begin{widetext}
	\begin{equation}
	B_1(R,T) dR= \frac{dR}{2} \int_{0}^{T- \frac{R}{v}} r_{{\rm eff}}(T_w) \lambda\left(\frac{R+v(T-T_w)}{2}\right) dT_w 
	\end{equation}
	where we use the fact that $dL=dR/2$. We can then evaluate $B_1(R,T)$ in the rescaled variable $r=R/(vT)$:
	\begin{equation}
	{\tilde B}_1(r,T)=
	\begin{dcases}
	\frac{1}{T^{\frac{1+\alpha+\alpha^2}{1+\alpha}}} \frac{2^\alpha \alpha L_0^\alpha}{ v^{1+\alpha} \tau_0^{{\alpha \over 1+\alpha}} } \int_{0}^{1-r} \frac{dt_w}{t_w^{{1 \over 1+\alpha}} (1+r-t_w)^{1+\alpha}}=\frac{1}{T^{\frac{1+\alpha+\alpha^2}{1+\alpha}}} f_{1,\alpha}(r) & \mathrm{if}\ 0<\alpha<1 \\
	\frac{1}{T^{\frac{1}{2}+\alpha}} \frac{2^\alpha \alpha L_0^\alpha}{  v^{1+\alpha} \tau_0^{{1 \over 2}}} \int_{0}^{1-r} \frac{dt_w}{t_w^{{1 \over 2}} (1+r-t_w)^{1+\alpha}}=\frac{1}{T^{\frac{1}{2}+\alpha}} f_{1,\alpha}(r)  & \mathrm{if}\ 1 \leq \alpha  
	\label{g1}
	\end{dcases},
	\end{equation}
	where we introduced the integration variable $t_w=T/T_w$.
	
	If $L<v(T-T_w)/2$ and $v(T-T_w)/3<L$ the motion displays two reflections and the final position satisfies the equation: $2L+R=v(T-T_w)$, so that $v(T-T_w)/3<(v(T-T_w)-R)/2<v(T-T_w)/2$. The second inequality is trivial, while the first gives: $T_w<T-3R/v$. The inequality cannot be satisfied if $R>vT/3$ indeed this process do not give contributions to distances larger than $vT/3$. Taking into account that $dL=dR/2$ we calculate the contribution of the process with 2 reflections obtaining
	in the rescaled variables:
	
	\begin{equation}
	{\tilde B}_2(r,T)=
	\begin{dcases}
	\frac{1}{T^{\frac{1+\alpha+\alpha^2}{1+\alpha}}} \frac{2^\alpha \alpha L_0^\alpha}{ v^{1+\alpha} \tau_0^{{\alpha \over 1+\alpha}}} \theta(1-3r) \int_{0}^{1-3r}\frac{dt_w}{t_w^{{1 \over 1+\alpha}} (1-r-t_w)^{1+\alpha} }
	=\frac{1}{T^{\frac{1+\alpha+\alpha^2}{1+\alpha}}} f_{2,\alpha}(r) & \mathrm{if}\ 0<\alpha<1                 \\
	\frac{1}{T^{\frac{1}{2}+\alpha}} \frac{2^\alpha \alpha L_0^\alpha}{ v^{1+\alpha} \tau_0^{{1 \over 2}}} \theta(1-3r) \int_{0}^{1-3r} \frac{{ dt_w}}{t_w^{{1 \over 2}} (1-r-t_w)^{1+\alpha}}=
	\frac{1}{T^{\frac{1}{2}+\alpha}} f_{2,\alpha}(r)      & \mathrm{if}\ 1 \leq \alpha  
	\label{g2}
	\end{dcases}
	\end{equation}
\end{widetext}
	where the $\theta$ function take into account that this process do not gives contributions for $R>vT/3$ .
	
	The case with a generic number $n$ of reflections occurs if $v(T-T_w)/(n+1)<L<v(T-T_w)/n$. For $n$ even we have $n/2$ reflections in $R=L$ and $n/2$ in $R=0$. Then $nL+R=v(T-T_w)$, $L=(v(T-T_w)-R)/n$, $T_w<T-(n+1)R/v$ and $R<vT/(n+1)$. 
	Taking into account that $dL=dR/n$ one can evaluate the contribution of this process obtaining for $f_{n,\alpha}(r)$ the result in Eq. (\ref{gndr}).
	For odd $n$ we have $(n+1)/2$ reflections at $R=L$ and $(n-1)/2$ reflection in $R=0$. Then $(n+1)L-R=v(T-T_w)$, $L=(v(T-T_w)-R)/(n+1)$, $T_w<T-nR/v$ and $dL=dR/(n+1)$. In this case we obtain Eq. (\ref{gnpr}). 
We can now sum all the contributions recovering Eq. (\ref{gtot}).

Let us analyze the behavior of $I_{\alpha}(r)$ at small $r$. First we notice that for $r\sim 0$ the integrals in Eqs. (\ref{gnpr},\ref{gndr}) display the following behavior:
\begin{eqnarray}
\int_{0}^{1-(n+1)r} \frac{dt_w}{t_w^{{1 \over 1+\alpha}} (1-r-t_w)^{1+\alpha} } & \sim &\nonumber\\ \int_{0}^{1-(n+1)r} \frac{dt_w}{t_w^{{1 \over 2}} (1-r-t_w)^{1+\alpha}} & \sim & \frac{1}{\alpha (nr)^\alpha } 
\end{eqnarray}
for even $n$, while for odd $n$
\begin{eqnarray}
\int_{0}^{1-nr} \frac{dt_w}{t_w^{{1 \over 1+\alpha}} (1+r-t_w)^{1+\alpha}} & \sim & \nonumber\\ \int_{0}^{1-nr} \frac{dt_w}{t_w^{{1 \over 2}} (1+r-t_w)^{1+\alpha}} & \sim&  \frac{1}{\alpha ((n+1)r)^\alpha } .
\end{eqnarray}
So that for small $r$:
\begin{equation}
f_{n,\alpha}(r)\sim
\begin{dcases}
\frac{\theta(1-(n+1)r)}{r^\alpha }  & \mathrm{if}\ n \ \mathrm{is \ even}                 \\
\frac{ \theta(1-nr)}{r^\alpha }    & \mathrm{if}\ n \ \mathrm{is \ odd}                 
\label{gnr1}
\end{dcases}
\end{equation}
Letting $r \to 0$ in Eq. (\ref{f0}) we get for $f_{0,\alpha}(r)$ the same expression of Eq. (\ref{gnr1})  ($n$ even). Summing  over $n$ we get $I_\alpha(r)\sim r^{-\alpha} {\mathcal{I}}(r)$;
where ${\mathcal{I} }(r)$ represents the largest odd integer smaller than $r^{-1}$. Since for small $r$, $\mathcal{I}(r)\sim r^{-1}$, we get $I_\alpha(r)\sim r^{-(1+\alpha)}$ and hence
\begin{equation}
{\tilde B}(r,T) \sim
\begin{dcases}
T^{-{1+\alpha+\alpha^2 \over 1+\alpha}}r^{-(1+\alpha)} & \mathrm{if}\ 0<\alpha<1 \\
T^{-{1 \over 2} -\alpha}r^{-(1+\alpha)}  & \mathrm{if}\ 1 \leq \alpha  
\label{old}
\end{dcases}
\end{equation}
which is the same equation obtained in \cite{levyrand} using a simple heuristic argument.
Our calculation shows that the $r^{-(1+\alpha)}$ behavior at small $r$ is given by two factors: the infinite density of a single reflection process diverges at small $r$ as $r^{-\alpha}$, but the number of processes (reflections) arriving in $r$ grows as $r^{-1}$ for $r\to 0$. This means that the density gets smoother close to the small $r$ limit, which is totally expected since it needs to match the smooth bulk statistics.

\section{Correlated L\'evy walks in the short time regime}
\label{appendixB}

Let us call $Q(R,T)$ the probability of making a jump at position $R$ and time $T$ and extracting at $T$ a new length from $\lambda(L)$.
One can write:
\begin{eqnarray}
Q(R,T) & = & \delta(R)\delta(T)+\frac{(1-p)}{2}\int \left[Q(R-L,T-L/v) \right. \nonumber \\ 
		&\ &\ \ \left. +Q(R+L,T-L/v) \right]\lambda(L) dL +\nonumber \\
	& + & {(1-p)} p \int Q(R,T-2L/V) \lambda(L) dL + \nonumber \\
	& + & \frac{(1-p)}{2} p^2 \int \left[Q(R-L,T-3L/v)\nonumber\right. \\
	& &\ \  \left.+Q(R+L,T-3L/v) \right]\lambda(L) dL \nonumber +\\
	& + & {(1-p)} p^3 \int Q(R,T-4L/V) \lambda(L) dL + \dots\nonumber \\
\label{ap1}
\end{eqnarray}
In the first term of the second member, the $\delta$-function takes into account that at time $T=0$ the walker is in $R=0$ 
and it makes a step choosing a new step length.
The second term represents processes where a new step length is extracted immediately before $T$ without any reflection;
the third term represents events where the walker makes a reflection before $T$ notice that in this case the length extraction 
occurs exactly in $R$ since in two steps the walker returns to the starting point; the fourth term represent events where the walker makes two reflections and so on.

Now we can sum over all the possible scattering events obtaining:
\begin{eqnarray}
& & Q(R,T)  =  \delta(R)\delta(T) +\nonumber\\
	& & + \frac{(1-p)}{2} \sum_n^\infty p^{2n} \int [Q(R-L,T-(2n+1)L/v)\nonumber\\
& &\ \ \ \ 	+Q(R+L,T-(2n+1)L/v) ]\lambda(L) dL + \nonumber \\
& & + {(1-p)} \sum_n^\infty p^{2n+1} \int Q(R,T-2(n+1)L/V) \lambda(L) dL.\nonumber \\
\label{ap2}
\end{eqnarray}
The probability $P(R,T)$ can be reconstructed from $Q(R,T)$ taking into account that a walker can arrive in $R$ only with a step of length $L'>R-L$ where $R-L$ is the position where $L'$ have been extracted form $\lambda(L')$. We have:
\begin{widetext}
\begin{eqnarray}
P(R,T) & = & \sum_n^\infty p^{2n} \int dL \left[Q(R-L,T-L/v-2nL'/v)+Q(R+L,T-L/v-2nL'/v) \right]\int_{L}^{\infty} dL' \lambda(L')  + \nonumber \\
& + &  \sum_n^\infty p^{2n+1} \int dL \left[Q(R-L,T+L/v-2nL'/v)+Q(R+L,T+L/v-2nL'/v) \right]\int_{L}^{\infty} dL' \lambda(L').
\label{ap2b}
\end{eqnarray}	
\end{widetext}

The first sum represents all processes performing an even number $2n$ of reflections between the extraction of the step length $L'$ and time $T$; while the second term 
describes the events where an odd number $2n+1$ of reflections occurs.
$T-L/v-2nL'/v$ and $T+L/v-2nL'/v$ are the starting times for getting in $R$ at time $T$ after $2n$ or $2n+1$ reflections respectively.

Let us consider $\tilde Q(k,\omega)$, i.e. the Fourier transform with respect $R$ and $T$ of $Q(R,T)$. From Eq. (\ref{ap2}) we get:
\begin{eqnarray}
& \tilde Q(k,\omega) &  = 1 + \nonumber\\
& & \tilde Q(k,\omega) \frac{(1-p)}{2} \sum_n^\infty p^{2n}  \left[\tilde \lambda  (\omega(2n+1)/v+k)\right.\nonumber \\
& &  \left.+\tilde \lambda (\omega(2n+1)/v-k)\right]+ \nonumber \\
& &  + \tilde Q(k,\omega) {(1-p)} \sum_n^\infty p^{2n+1} \tilde \lambda (\omega(2n+2)/v) .\nonumber\\
\label{ap3}
\end{eqnarray}
where $\tilde \lambda(\cdot)$ is the Fourier transform of $\lambda(\cdot)$. 
Now we can expand $\tilde \lambda(\cdot)$ for small $\omega$ and $k$; keeping only the leading terms in Eq. (\ref{ap3}) for $\alpha>2$ we have:
\begin{eqnarray}
& &\tilde Q(k,\omega)  =  1 + \tilde Q(k,\omega) {(1-p)}\cdot \nonumber\\
& & \ \ \ \ \ \  \cdot\sum_n^\infty p^{2n}  \left[1+(2n+1)i \omega \langle L \rangle /v -k^2 \langle L^2\rangle/2 \right]+ \nonumber \\
& & +  \tilde Q(k,\omega) {(1-p)} \sum_n^\infty p^{2n+1} \left[1+(2n+2)i \omega \langle L \rangle/v \right].\nonumber\\
\label{ap4}
\end{eqnarray}
Summing over $n$ we have
\begin{equation}
\tilde Q(k,\omega) =  1 + \tilde Q(k,\omega) \left(1+ \frac{i \omega\langle L\rangle}{(1-p)v} - \frac{k^2 \langle L^2\rangle}{2(1+p)}\right)
\label{ap5}
\end{equation}
i.e.:
\begin{equation}
\tilde Q(k,\omega) =  \frac{1}{\frac{k^2 \langle L^2\rangle}{2(1+p)}-\frac{i \omega\langle L\rangle}{(1-p)v}}.
\label{ap6}
\end{equation}
For $1<\alpha<2$ we get:
\begin{eqnarray}
& & \tilde Q(k,\omega)  =  1 + \tilde Q(k,\omega) {(1-p)} \cdot \nonumber\\
& &\ \ \ \ \cdot\sum_n^\infty p^{2n}  \left[1+(2n+1)i \omega \langle L \rangle /v -|k|^\alpha C_\alpha L_0^\alpha \right]+ \nonumber \\
& & +   \tilde Q(k,\omega) {(1-p)} \sum_n^\infty p^{2n+1} \left[1+(2n+2)i \omega \langle L \rangle/v \right].\nonumber\\
\label{ap7}
\end{eqnarray}
where $C_\alpha$ is a number depending only on $\alpha$ and $L_0$ is the cut-off in Eq. (\ref{lambdaL}). Then summing we have:
\begin{equation}
\tilde Q(k,\omega) =  \frac{1}{\frac{k^\alpha L_0^\alpha C_\alpha}{(1+p)}-\frac{i \omega\langle L\rangle}{(1-p)v}}.
\label{ap8}
\end{equation}
Fourier transforming Eq.(\ref{ap2b}) after some algebra one obtain that for $\alpha>1$ we have 
$\tilde P(k,\omega)=\tilde Q(k,\omega)\langle L\rangle /(v(1-p))$, so that  for $\alpha>2$
\begin{equation}
\tilde P(k,\omega) =  \frac{\frac{\langle L\rangle}{(1-p)v} }{\frac{k^2 \langle L^2\rangle}{2(1+p)}-\frac{i \omega\langle L\rangle}{(1-p)v}}.
\label{ap6b}
\end{equation}
and for $1<\alpha<2$
\begin{equation}
\tilde P(k,\omega) =  \frac{\frac{\langle L\rangle}{(1-p)v}}{\frac{k^\alpha L_0^\alpha C_\alpha}{(1+p)}-\frac{i \omega\langle L\rangle}{(1-p)v}}.
\label{ap8b}
\end{equation}
From Eq.s (\ref{ap6b},\ref{ap8b}) we immediately have that introducing the new variables
$\omega'=\omega/(1-p)$, $k'=k/(1+p)^{1/2}$ for $\alpha>2$ and $\omega'=\omega/(1-p)$, $k'=k/(1+p)^{1/\alpha}$ for $1<\alpha<2$ we obtain
the standard PDF functions for a L\'evy walks. In particular, we get a Gaussian scaling function for $\alpha>2$ and a L\'evy-like scaling function depending only on $\alpha$ for $1<\alpha<2$. In this framework, we can introduce the scaling length  
$\ell(T)\sim ((1-p)/(1+ p)T)^{1/2}$ and $\ell(T)\sim ((1-p)/(1+p)T)^{1/\alpha}$ for $\alpha>2$ and $1<\alpha<2$ respectively; in this way, we obtain a perfect rescaling of the PDF for different values of the parameter $p$ as shown in Fig. \ref{scalSHc} ($\alpha=1.7$).

Let us finally consider the case $\alpha<1$; if we expand $\tilde Q(k,\omega)$ at small $k$ and $\omega$,  we get:
\begin{widetext}
\begin{eqnarray}
 \tilde Q(k,\omega) & = & 1 + \tilde Q(k,\omega) {(1-p)} \sum_n^\infty p^{2n} \left[1+ \left(|(2n+1)\omega/v-k|^\alpha+|(2n+1)\omega/v+k|^\alpha\right) C_\alpha L_0^\alpha \right]+ \nonumber \\
& + &  \tilde Q(k,\omega) {(1-p)} \sum_n^\infty p^{2n+1} \left[1+|(2n+1)\omega/v|^\alpha
C_\alpha L_0^\alpha \right].
\label{ap9}
\end{eqnarray}
\end{widetext}
Where $C_\alpha$ are suitable complex coefficients.  Since $\omega$ and $k$ always appears in a linear combination
linear ballistic relation between space and time is expected in this case. However, a simple summation of the different
terms corresponding to different $n$ is not possible and a scaling function which depends non trivially 
on $p$ is clearly expected from Eq. (\ref{ap9}). Moreover in this case also the relation between $\tilde P(k,\omega)$ and 
$\tilde Q(k,\omega)$ is not a simple proportionality.


\begin{thebibliography}{}
\bibitem{note1}
It should be noted that the number of zero crossing is in fact infinite, since the paths are continuous, but still we claim we may consider the biggest jump. In practice, in simulations one has a finite $\Delta t$ element, so one can check that while the size of the jump is decreasing,  the number of zero crossing is increasing, the distribution of the biggest jump is stable with respect to the discretization of time.
\end{thebibliography}
\end{document}